\documentclass[twocolumn, twocolappendix]{aastex701}
\usepackage{bm}

\newcommand{\mytilde}{\raise.19ex\hbox{$\scriptstyle\sim$}}

\usepackage[utf8]{inputenc}
\usepackage{CJK}
\usepackage{amsmath}
\usepackage{lineno}
\usepackage{tabularx}
\usepackage{booktabs}
\usepackage{multirow}
\usepackage{threeparttable}

\setcitestyle{notesep={; }}

\received{}
\revised{}
\accepted{}

\submitjournal{}

\shorttitle{WL Filament Analysis of A2744}
\shortauthors{Cha et al. }

\begin{document}

\title{Weak-lensing Analysis of Intracluster Filaments in Abell 2744: Matched-filter Scans and Stepwise 2D Tracing}

\correspondingauthor{M. James Jee}

\author[0000-0001-7148-6915]{Sangjun Cha}
\affiliation{Department of Astronomy, Yonsei University, 50 Yonsei-ro, Seoul 03722, Korea}
\email[show]{sang6199@yonsei.ac.kr}

\author[0000-0002-4462-0709]{Kyle Finner}
\affiliation{IPAC, California Institute of Technology, 1200 E California Blvd., Pasadena, CA 91125, USA}
\email[]{kfinner@ipac.caltech.edu}

\author[0000-0002-5751-3697]{M. James Jee}
\affiliation{Department of Astronomy, Yonsei University, 50 Yonsei-ro, Seoul 03722, Korea}
\affiliation{Department of Physics and Astronomy, University of California, Davis, One Shields Avenue, Davis, CA 95616, USA}
\email[show]{mkjee@yonsei.ac.kr} 

\author[0000-0002-5688-0663]{Andrea Grazian}
\affiliation{INAF--Osservatorio Astronomico di Padova, Vicolo dell'Osservatorio 5, I-35122, Padova, Italy}
\email[]{andrea.grazian@inaf.it}

\begin{abstract}

We present a weak-lensing (WL) analysis of filamentary structures in the merging galaxy cluster Abell 2744 using wide-field Magellan/MegaCam imaging data. 
We employ two complementary techniques: standard matched-filter scans to identify global orientations, and a new stepwise 2D tracing method to reconstruct locally varying filament orientations. The matched-filter analysis detects coherent filamentary features in the northwest and east directions across both inner ($1.0\text{--}2.2$~Mpc) and outer ($2.2\text{--}3.4$~Mpc) annuli. 
However, while the northwest filament yields consistent constraints across both regions, parameter inference for the eastern structure remains unstable and radially inconsistent when using the fiducial reference point. Additional matched-filter scans centered on the three lensing mass peaks show that the inferred directions depend on the adopted reference point,
We demonstrate that re-characterizing the eastern structure using the locally preferred stepwise-traced direction significantly resolves these tensions, improving fit quality and bringing inner and outer constraints into agreement. These results show that stepwise 2D tracing complements global matched-filter scans by tracing locally coherent directions without requiring the full structure to pass through a single predefined point.
Furthermore, the detected filaments align well with diffuse X-ray structures and previously identified merger axes, supporting their physical connection to the cluster's mass assembly.
\end{abstract}

\section{Introduction} 
In the $\Lambda$CDM standard cosmology, structure grows hierarchically, and the matter distribution is organized into a cosmic web of voids, filaments, and nodes \citep{1996Natur.380..603B, 2006Natur.440.1137S}. Galaxy clusters form at the nodes, and connected filaments provide pathways for mass assembly \citep[e.g.,][]{2014MNRAS.441.2923C, 2021MNRAS.502..714R, 2022A&A...661A.115G}. Such filamentary accretion contributes to cluster growth and is linked to diverse dynamical processes in and around clusters \citep[e.g.,][]{2013SSRv..177..195R, 2019SSRv..215....7W, 2020MNRAS.494.5473K}. 

Filaments have been studied extensively as key components of the cosmic web. In observations, filaments have been traced with galaxy distributions from spectroscopy \citep[e.g.,][]{2004ApJ...609L..49E, 2009MNRAS.399..683J, 2011ApJ...736...59Z, 2014ApJS..213...35G, 2016A&A...588A..69D} and with X-ray and SZ measurements of the hot gas \citep[e.g.,][]{2008A&A...482L..29W, 2017A&A...606A...1A, 2022MNRAS.509.1109M, 2022MNRAS.510.3335H, 2025A&A...698A.270M}. Despite these efforts, robust detections and measurements remain challenging because filaments have low projected mass density. Furthermore, these baryonic tracers have limitations. Galaxy searches are limited by sample selection, and X-ray/SZ measurements are sensitive to gas physics and the dynamical state of the system.

Weak lensing (WL) offers a unique probe of filamentary structures, capable of mapping the mass distribution without requiring dynamical assumptions. Given the low density contrast of the filamentary structures, previous WL analyses have mainly focused on stacked filaments \citep[e.g.,][]{2016MNRAS.457.2391C, 2017MNRAS.468.2605E, 2020MNRAS.495.3695K, 2020A&A...633A..89X}. Only a few studies have reported filament detections through WL analyses of individual clusters using the mass map and the matched-filter technique \citep[e.g.,][]{2012Natur.487..202D, 2012MNRAS.426.3369J, 2024NatAs...8..377H, 2025arXiv251026318S}.

In this study, we present a WL analysis of intracluster filaments in Abell 2744 using deep, wide-field Magellan/MegaCam images ($\sim26\arcmin\times26\arcmin$). Abell 2744 is a massive and dynamically complex merging galaxy cluster that has been widely studied \citep[e.g.,][]{2011MNRAS.417..333M, 2011ApJ...728...27O, 2013A&A...551A..24V, 2016ApJ...817...24M, 2017ApJ...845...81P, 2021A&A...654A..41R, 2023ApJ...951..140C, 2023ApJ...952...84B, 2024ApJ...961..186C, 2024A&A...684A.193A, 2025ApJS..277...28F, 2025arXiv251027291W}. In particular, X-ray surface brightness observations have been used to investigate diffuse large-scale structures and filament candidates around this system \citep[][]{2015Natur.528..105E, 2024A&A...692A.200G}. However, in such a complex merger, diffuse X-ray features can be influenced by baryonic physics and the thermodynamic state of the gas, complicating the interpretation. \citet{2015Natur.528..105E} also incorporated WL constraints, but their analysis relied on a relatively low background WL source density. Given the intrinsically low density contrast of intracluster filaments, a high source density is essential to suppress shape noise and robustly detect such faint structures.

Even with deep imaging data, defining the filament orientation remains a challenge. WL mass maps provide an intuitive view of the projected mass distribution \citep[e.g.,][]{2016ApJ...817...24M, 2017ApJ...851...46F, 2024ApJ...973...79A, 2026ApJ...999L...1S} and make it straightforward to trace elongated features in two dimensions. However, their fidelity depends on the chosen reconstruction algorithm, smoothing scale, and regularization scheme. On the other hand, the matched-filter technique adopts an optimal filter to maximize the expected S/N for a given filament model \citep[][]{2005A&A...442..851M, 2013A&A...559A.112M}. 
While powerful, this method represents each candidate with a straight axis passing through a predefined reference point. The inferred direction can depend on the adopted reference point, and a single global orientation may not capture locally varying directions or curved morphologies.

To address these limitations, we introduce a comprehensive WL framework that synergizes matched-filter scans with our new stepwise 2D tracing method, demonstrated on this merging system. 
While the matched-filter efficiently identifies global filament orientations, 
the stepwise tracing maps locally coherent directions without requiring the full structure to pass through a single predefined point. It also retains the flexibility to follow locally varying directions and curved morphologies that are not well represented by a single global orientation.
We also explicitly compare our WL results to the X-ray morphology and merger geometry reported in the literature.

In Section~\ref{section:data}, we introduce the data and reduction process. Section~\ref{section:method} describes the WL filament analysis method. We present our results in Section~\ref{section:result} and discuss them in Section~\ref{section:discussion}. We conclude in Section~\ref{section:conclusion}.
Unless stated otherwise, we assume a flat $\Lambda$CDM cosmology with the matter density $\Omega_{M}=1-\Omega_{\Lambda}=0.3$ and the dimensionless Hubble constant parameter $h=0.7$. At the redshift of Abell 2744 ($z=0.308$), the plate scale is $0.272 ~\rm Mpc ~\rm arcmin^{-1}$.

\section{Data}\label{section:data}
\subsection{Magellan Images}
Observations of Abell 2744 were obtained with MegaCam \citep{2015PASP..127..366M}  on the Magellan 2 Clay Telescope (6.5m). 
MegaCam has an array of 36 2k × 4k CCD detectors with a $\sim25\arcmin×25\arcmin$ field of view per exposure.
The detectors have a plate scale of 0\farcs08, which provides sufficient sampling when seeing is optimal. 
The observations were carried out on 2018 September 8-9 (PI L. Abramson) under prime seeing conditions with average seeing values in the range of 0\farcs5-0\farcs7 \citep{2022ApJ...938L..14M}. Images were obtained in the $g$-, $r$-, and $i$-band filters. We reduced the data with custom scripts utilizing the calibration images (bias, darks, sky flats) taken during the same observing run as the science frames. The MegaCam $i$-band imaging contains significant fringing. Fringing was removed by constructing a fringe template for each detector. The templates for each detector were generated by determining the median frame after masking objects and sky normalizing the science exposures. For each science frame, this template was scaled by minimizing the residual background scatter and then subtracted to remove the fringe signal. 

The calibrated frames were provided to SExtractor \citep{1996A&AS..117..393B} to create object catalogs for alignment. Image registration and a refined distortion correction were achieved with SCAMP \citep{bertinscamp}. Finally, we stacked the component frames into a filter-dependent mosaic image with SWARP \citep{bertinswarp}. The final mosaic image used for the WL analysis covers $\sim26\arcmin\times26\arcmin$.

\subsection{WL Data}
\subsubsection{PSF Modeling and Shape Measurement}
\begin{figure}
\centering
\includegraphics[width = \columnwidth]{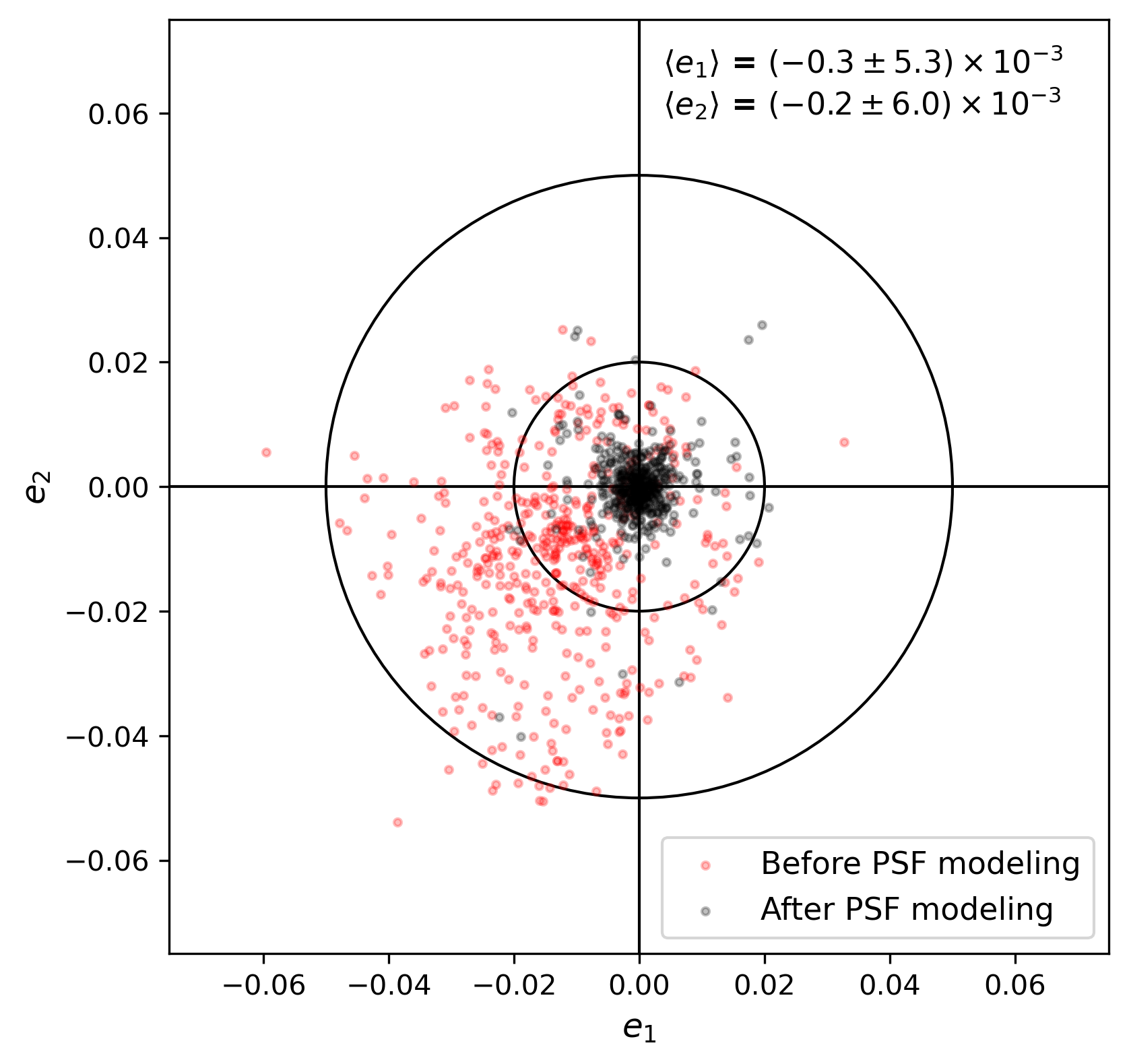}
\caption{PSF correction for Abell 2744. The red dots mark the ellipticities of stars before correction. The black dots indicate the corrected ellipticities after PSF correction. The average residuals of the ellipticity components show the robustness of our PSF modeling.}
\label{fig:psf_residual}
\end{figure}

Accurate WL shape measurements require a precise modeling of the point-spread function (PSF), since spatial variations in PSF anisotropy and size can introduce spurious ellipticity patterns or dilute the shear signal. 
We modeled the spatially varying PSF using a principal component analysis (PCA) of stellar postage stamps \citep{2007PASP..119.1403J, 2011PASP..123..596J}. A clean sample of unsaturated, isolated, high-quality stars was selected from size–magnitude relations. 
Stellar centroids were determined from the {\tt SExtractor} windowed centroid coordinates, and each stellar image was recentered according to its subpixel offset from the nearest pixel center using a third-order B-spline interpolation kernel before constructing a fixed-size stamp.
For each CCD frame, we subtracted the mean PSF from the stellar images and performed PCA on the residuals, retaining the first 21 principal components.  
This choice accounts for more than 99\% of the total variance for most CCDs in our data and has been commonly adopted in previous WL studies \citep[e.g.,][]{2007PASP..119.1403J, 2024ApJ...962..100H, 2025ApJS..277...28F}.
The PCA coefficients of the individual stars were determined from this basis.
To obtain a continuous PSF model within each CCD, we fitted third-order polynomials to the spatial dependence of each PCA coefficient and reconstructed the PSF at arbitrary positions by adding the interpolated PCA residuals back to the mean PSF. For the mosaic image, we generated PSF models at the object position on a frame-by-frame basis and then combined them into an effective coadded PSF using the relevant frame weights and orientations \citep[see][]{2015ApJ...802...46J, 2017ApJ...851...46F, 2024ApJ...962..100H, 2025ApJS..277...28F}. We evaluated the PSF model performance by comparing the ellipticities of observed stars to those predicted by the PSF model using the quadrupole moments as shown in \citet{2007PASP..119.1403J}. In Figure~\ref{fig:psf_residual}, we show the PSF corrections for Abell 2744.

Galaxy shapes were measured with a forward modeling approach, in which an elliptical Gaussian surface-brightness model is convolved with the local PSF and fitted to each galaxy postage stamp image. A corresponding rms noise stamp was used to weight pixels, and nearby contaminants were masked using the {\tt SExtractor} segmentation map when necessary. We determined the best-fit parameters by minimizing
\begin{equation}
\chi^2=\sum\frac{\left[I-\left(G\otimes P\right)\right]^2}{\sigma_{\rm rms}^2},
\end{equation}
where $I$ is the observed postage-stamp image, $P$ is the PSF model evaluated at the object position, and $G$ is the elliptical Gaussian model. The fit was performed with a Levenberg–Marquardt least-squares solver \citep[{\tt MPFIT};][]{2009ASPC..411..251M}. Following the standard implementation in our WL pipeline, we fixed the centroid and background to the {\tt SExtractor} values and fit the remaining parameters, from which the complex 
ellipticity components were computed as
\begin{equation}
e_1=\frac{a-b}{a+b}\cos 2\phi,\quad
e_2=\frac{a-b}{a+b}\sin 2\phi,
\end{equation}
with semi-major/semi-minor axes a, b, and position angle $\phi$. Model fitting is subject to shear measurement biases (e.g., model bias, noise bias). We therefore applied a multiplicative calibration factor to account for the net dilution of the measured ellipticities. In this study, we adopt $m=1.14$ from \citet{2024ApJ...973...79A}, who analyzed Magellan/MegaCam data using essentially the same reduction and shape measurement framework as adopted in this study. The calibration factor was derived from image simulations matched to the observations by comparing the input and recovered shears. The resulting additive bias was negligible. For more details, we refer readers to \citet{2011PASP..123..596J}, \citet{2013ApJ...765...74J}, and \citet{2015MNRAS.450.2963M}.

\subsubsection{Source Selection and Redshift Estimation}\label{background_source_redshift}
\begin{figure}
\centering
\includegraphics[width = \columnwidth]{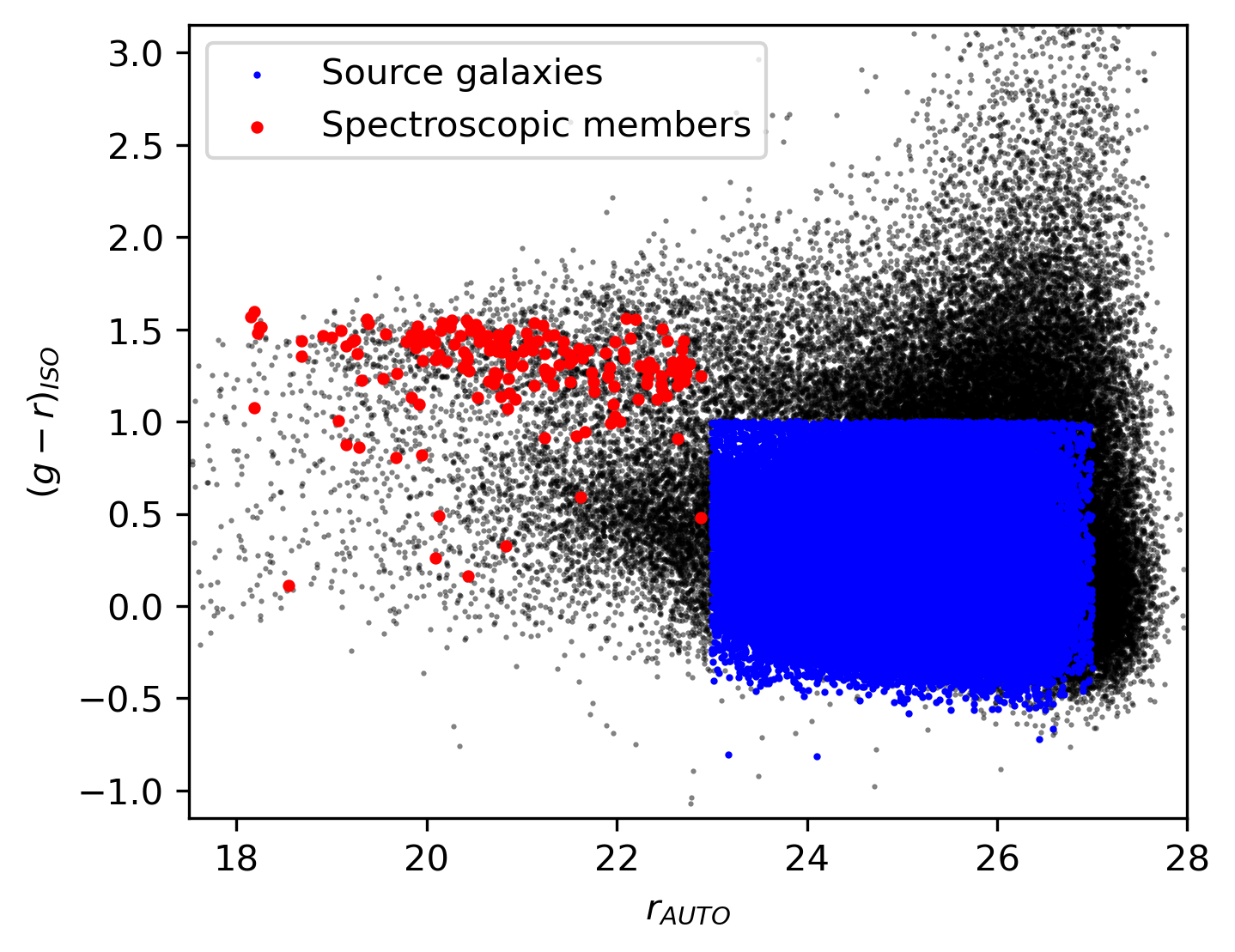}
\caption{Color--magnitude relation for Abell 2744 using $r$ [MAG\_AUTO] and $(g-r)$ [MAG\_ISO] from {\tt SExtractor}. The red dots indicate spectroscopically confirmed cluster members from \citet{2023ApJ...952...84B} with $|z-z_{\rm cl}|<0.02$ where $z_{\rm cl}=0.308$. Background sources selected for the WL analysis are marked as blue circles (see \S\ref{background_source_redshift}).}
\label{fig:CMD}
\end{figure}

We selected background sources for the WL analysis using color-magnitude cuts. 
Given the cluster redshift of Abell 2744 and the location of the 4000\AA\ break, the red sequence is well identified with the $g$ and $r$ filters. In this study, we selected sources with $23<r<27$ and $-1<g-r<1$ as the magnitude and color cuts, respectively (see Figure~\ref{fig:CMD}). 
Although foreground contamination is a potential concern, recent studies suggest its impact is not significant. For example, \citet{2025ApJS..277...28F} report the foreground contamination fraction to be at most $\sim10\%$ at $z_{\rm lens}=0.3$, even with a looser cut of $g-r<2$.
After this selection, we removed sources that do not satisfy the following quality cuts: (1) successful model fitting ({\tt MPFIT} STATUS$=1$), (2) semi-minor axis $>0.3$ pixels and semi-major axis $<30$ pixels, (3) total ellipticity $e<0.9$, (4) ellipticity measurement error $<0.3$, and (5) {\tt SExtractor} flag $<4$. 
The final catalog contains 48,870 galaxies, corresponding to a mean source density of $\sim66~{\rm arcmin^{-2}}$.

In gravitational lensing, the shear amplitude depends on the lens and source redshifts. 
Since we do not have redshifts for individual sources, we estimated an effective lensing efficiency using the COSMOS2020 photometric redshift catalog \citep{2022ApJS..258...11W}. 
Because the COSMOS2020 catalog does not provide Magellan/MegaCam photometry, we used the Subaru/Hyper Suprime-Cam $g$ and $r$ magnitudes as proxies for our $g$ and $r$ bands, following \citet{2024ApJ...973...79A}. 
To compute the effective lensing efficiency, we applied the same magnitude and color cuts to COSMOS2020 as used for our source selection. To account for depth differences, we weighted the COSMOS2020 galaxies by the ratio of the observed number densities in the Abell 2744 and COSMOS fields. 
The effective $\beta$ is computed as follows:
\begin{equation}
\langle\beta\rangle=\left\langle \max\left(0,\frac{D_{\rm ds}}{D_{\rm s}}\right)\right\rangle,
\end{equation}
where $D_{\rm s}$ and $D_{\rm ds}$ are the angular diameter distances to the source and between the lens and the source, respectively.
We obtained $\langle\beta\rangle=0.638$, which corresponds to an effective source redshift of $z_{\rm eff}=1.03$. We note that the source redshift distribution inferred from the finite field may be affected by cosmic variance. The resulting uncertainty in the mean lensing efficiency may affect the fitted filament parameters and S/N. However, because this effect is expected to act similarly across all position angles, the preferred filament directions and stepwise morphology are expected to be relatively insensitive to this uncertainty.

\subsection{Cluster Member Galaxies}
To compare our filament candidates with the cluster member galaxy distributions, we used the spectroscopic redshift catalog compiled by \citet{2011ApJ...728...27O}. This catalog combines AAOmega multi-object spectroscopy obtained with the 3.9 m Anglo-Australian Telescope and redshifts derived from VIMOS observations on VLT \citep{2009A&A...500..947B}.

We selected cluster member galaxies with $|z-z_{\rm cl}|<0.02$, where $z_{\rm cl}=0.308$. In addition, we applied a magnitude cut of $r_{\rm F}<20.5$, as adopted in \citet{2024A&A...692A.200G}. With this cut, the spectroscopic completeness is $\gtrsim90\%$ (see Fig.~9 of \citealt{2011ApJ...728...27O}). In total, we obtained 302 spectroscopically confirmed cluster members (located within $\sim15\arcmin$ of the cluster center). The cluster member distribution is used exclusively for post-analysis comparison with our filament candidates and is not used for filament detection.

\section{Method}\label{section:method}
\subsection{WL Formalism}
We briefly summarize the WL formalism used in this section. For general reviews, we refer to \citet{bartelmann2001}, \citet{2011A&ARv..19...47K}, and \citet{hoekstra2013}.

In the WL regime, the lens mapping can be locally linearized, and the image-plane transformation is described by a Jacobian matrix $\mathbf{A}$:
\begin{equation}
\boldsymbol{\theta}' = \mathbf{A}\boldsymbol{\theta}, \quad
\mathbf{A} =
\begin{pmatrix}
1-\kappa - \gamma_1 & -\gamma_2 \\
-\gamma_2 & 1-\kappa + \gamma_1
\end{pmatrix},
\end{equation}
where $\kappa$ is the convergence and $\gamma_1$ and $\gamma_2$ are the two components of the shear.
The convergence is defined as the projected surface mass density in units of the critical surface density,
\begin{equation}
\kappa(\boldsymbol{\theta}) = \frac{\Sigma(\boldsymbol{\theta})}{\Sigma_{\rm c}}, \quad
\Sigma_{\rm c} = \frac{c^2}{4\pi G}\frac{D_{\rm s}}{D_{\rm d}D_{\rm ds}},
\end{equation}
where $D_{\rm d}$, $D_{\rm s}$, and $D_{\rm ds}$ are the angular diameter distances to the lens, to the source, and between the lens and the source, respectively.
The reduced shear is
\begin{equation}
g = \frac{\gamma}{1-\kappa},
\end{equation}
where $\gamma \equiv \gamma_1+i\gamma_2$.
In the WL regime ($\kappa\ll 1$), one has $g \simeq \gamma$.

Given a convergence field, the shear can be obtained from the lensing potential derivatives. Defining the lensing potential $\psi(\boldsymbol{\theta})$, the convergence and shear are given by
\begin{equation}
\kappa = \frac{1}{2}\left(\psi_{11}+\psi_{22}\right), \quad
\gamma_1 = \frac{1}{2}\left(\psi_{11}-\psi_{22}\right), \quad
\gamma_2 = \psi_{12},
\end{equation}
where $\psi_{ij}\equiv \partial^2\psi/\partial\theta_i\partial\theta_j$.
Equivalently, through a convolution relation,
\begin{equation}
\gamma(\boldsymbol{\theta}) = \frac{1}{\pi}\int d^2\boldsymbol{\theta}' \, D(\boldsymbol{\theta}-\boldsymbol{\theta}')\,\kappa(\boldsymbol{\theta}'),
\end{equation}
where $D$ is the complex convolution kernel
\begin{equation}
D(\boldsymbol{\theta}) = -\frac{1}{(\theta_1-i\theta_2)^2}.
\end{equation}

\subsection{WL Mass Reconstruction}
\begin{figure}
\centering
\includegraphics[width = \columnwidth]{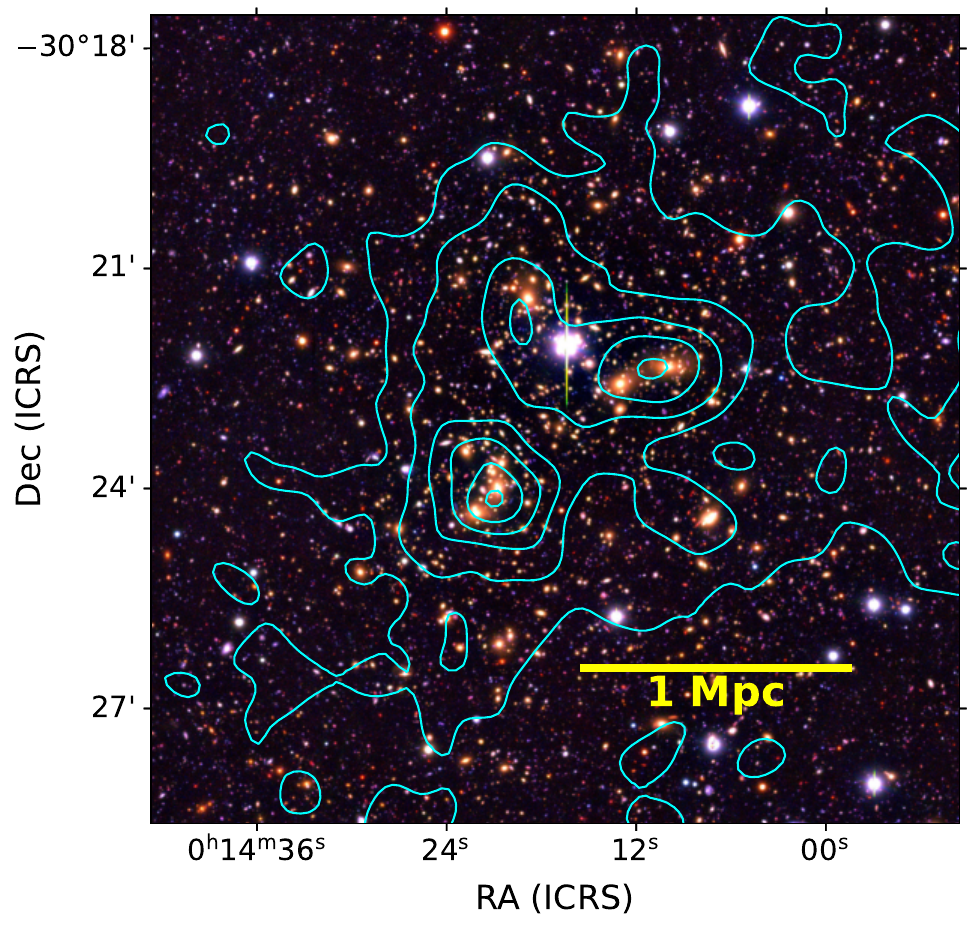}
\caption{Reconstructed mass map of the central $3~{\rm Mpc} \times 3~{\rm Mpc}$ region of Abell 2744. The cyan contours represent the reconstructed WL mass map at convergence levels of $\kappa = 0.12, 0.24, 0.36, 0.48, 0.6,$ and $0.7$. The background color-composite image is generated using Subaru/SC data, with the B, R, and $z$ bands assigned to the blue, green, and red channels, respectively \citep{2025ApJS..277...28F}.}
\label{fig:a2744_massmap_zoom}
\end{figure}

We used the free-form MAximum-entropy ReconStruction ({\tt MARS}) algorithm \citep[][]{2022ApJ...931..127C, 2023ApJ...951..140C} to reconstruct the WL mass map of Abell 2744. We minimized a target function consisting of a WL $\chi^2$ term and a regularization term. The WL $\chi^2$ term minimizes residuals between the observed and model-predicted reduced shear. The regularization term maximizes the cross-entropy between the $\kappa$ field and a prior, defined as the smoothed mass map from the previous epoch and updated at each iteration. This regularization enables {\tt MARS} to suppress overfitting and maintain smoothness \citep[e.g.,][]{2024ApJ...961..186C, 2025ApJ...987L..15C, 2025ApJ...979...13P, 2025ApJ...993..226S}. We refer to \citet{2024ApJ...961..186C} for details.

For the reconstruction, we used a $400\times400$ grid covering a $\sim10\,\mathrm{Mpc}\times10\,\mathrm{Mpc}$ region and initialized the convergence field with a flat $\kappa=0$ field. We allowed negative $\kappa$ values to mitigate regularization-induced positive bias in low-signal regions. To keep the entropy regularization well defined when $\kappa<0$, we followed the formalism of \citet{1998MNRAS.298..905H}, which has been adopted in previous lensing analyses \citep[e.g.,][]{2008ApJ...684..177U, 2026ApJ...997...18C}. In this study, the reconstructed WL mass map is used exclusively for visualization and qualitative comparison and is not utilized for the identification or detection of filamentary structures.

Figure~\ref{fig:a2744_massmap_zoom} displays the central $3~{\rm Mpc} \times 3~{\rm Mpc}$ region of the mass map reconstructed using {\tt MARS}. 
The recovered mass peaks spatially coincide with the locations of the BCGs, remarkably consistent with the substructures resolved in high-precision JWST lens models \citep[e.g.,][]{2024ApJ...961..186C}. This excellent agreement with the JWST results supports the fidelity of our WL measurement and systematics control.

\subsection{Identification of Filament Directions}\label{matched_filter_method}
We use the matched-filter technique to constrain filament orientations around Abell 2744. Designed to maximize the S/N value for a specific structural profile and noise model, this kernel-based approach has been employed in filament detection studies \citep[e.g.,][]{2005A&A...442..851M, 2013A&A...559A.112M, 2024NatAs...8..377H, 2025arXiv251026318S}.
Our analysis follows the same matched-filter framework as the previous studies. 
However, we construct the filter using inverse-variance weights for individual galaxies, which account for galaxy-dependent uncertainties in the WL catalog, but do not include correlations in the LSS noise covariance.

Since the global mass structure of Abell 2744 is characterized by three primary halos, we adopt their approximate geometric center as the fiducial reference point for our filament scans.
In Appendix~\ref{offset_node_cluster}, we repeat the matched-filter scan using each of the three lensing mass peaks as the reference point to test the sensitivity of the inferred directions to this choice.
For the matched-filter scan, we consider two radial ranges to probe possible scale dependence: an inner annulus of $1.0-2.2 \mathrm{~Mpc}$ and an outer annulus of $2.2-3.4 \mathrm{~Mpc}$. 
Due to the complex triangular substructure present within the inner $1 \mathrm{~Mpc}$, we exclude this region from our analysis.

To maximize sensitivity, we determine candidate filament orientations 
without subtracting the cluster contribution from the WL signal.
This follows the methodology of previous studies, which successfully identified orientations without halo subtraction and found them to be consistent with independent tracers \citep[][]{2024NatAs...8..377H, 2025arXiv251026318S}. 
However, to characterize the filament properties, we perform parameter inference on the halo-subtracted field, thereby isolating the target signal from the host cluster's contribution.

\subsubsection{Analytic Filament Profile}
In this study, we employ a projected surface mass density profile used in \citet{2013A&A...559A.112M}, which was derived from cosmological simulations \citep{2005MNRAS.359..272C, 2010MNRAS.401.2257M}. The analytic form of the filament is defined as:
\begin{equation}
    \kappa(h)=\frac{\kappa_0}{1 + (h/h_c)^2},
\label{filament_profile}
\end{equation}
where $\kappa_0$ is the maximum convergence. $h_c$ is a scale radius where the convergence drops to half of its peak value. In this model, the filament-induced shear is oriented perpendicular to the filament axis (i.e., $\gamma_+=\kappa(h)$ and $\gamma_\times=0$). Following \citet{2024ApJ...962..100H}, we refer to the $+$ and $\times$ components as the tangential and cross components, respectively.
If the filament axis passes through the origin, with orientation angle $\theta_{\rm f}$ measured counterclockwise from the x-axis, the shear is decomposed into the following components: 
\begin{align}
\begin{split}
    \gamma_{\rm +} = \gamma_1 {\rm cos}[2(\theta_{\rm f}-\pi/2)] + \gamma_2 {\rm sin}[2(\theta_{\rm f}-\pi/2)],\\
    \gamma_{\rm \times} = \gamma_1 {\rm cos}[2(\theta_{\rm f}-\pi/4)] + \gamma_2 {\rm sin}[2(\theta_{\rm f}-\pi/4)],
\label{filament_shear_component}
\end{split}
\end{align}
where $\gamma_1$ and $\gamma_2$ are the shear components of source galaxies.

\subsubsection{Matched-filter Statistic}\label{matched_filter_stat}
In this study, we follow the discrete matched-filter framework of \citet{2013A&A...559A.112M} and adopt the normalization used in \citet{2024NatAs...8..377H}. 
In addition, we incorporate galaxy-dependent uncertainties through inverse-variance weights assigned to individual galaxies.

In \citet{2024NatAs...8..377H}, the matched-filter statistic is given as
\begin{equation}\label{matched_filter_HH2024}
\Gamma(\theta)=\frac{1}{W(\theta)}\sum_i \gamma_i\,\Psi_i(\theta),
\quad
W(\theta)=\sum_i \Psi_i(\theta),
\end{equation}
where $\Psi_i(\theta)$ is the optimal filter value at the position of the $i$th galaxy and $\gamma_i$ denotes the tangential or cross component, and $W(\theta)$ is the normalization factor given by the sum of the filter values over all galaxies. In their implementation, the filter is constructed in Fourier space as
\begin{equation}
\hat{\Psi}_{\rm F}(\mathbf{k})=\hat{\tau}(\mathbf{k})/P_n(\mathbf{k}).
\label{filter_HH2024}
\end{equation}
Here, $\hat{\Psi}_{\rm F}(\mathbf{k})$ is the Fourier transform of the real-space filter profile, $\hat{\tau}(\mathbf{k})$ is the Fourier transform of the predicted filament shear, and $P_n(\mathbf{k})$ is the noise power spectrum (shape noise + LSS). 

In this study, we compute $\Gamma(\theta)$ using Equation~\ref{matched_filter_HH2024} with the following filter weights
\begin{equation}
\Psi_{i, \rm R}(\theta)=\psi_i(\theta)\,w_i,
\label{real_space_filter}
\end{equation}
where $\psi_i(\theta)$ is the predicted shear from the filament profile (Equation \ref{filament_profile}), and $w_i$ is the inverse-variance weight of the $i$th galaxy:
\begin{equation}
w_i=\frac{1}{\sigma^2_{i,{\rm base}}}
=\frac{1}{\sigma_{\rm shape}^2+\sigma_{i,{\rm meas}}^2+\sigma_{i,{\rm halo}}^2}.
\label{inverse_variance_weight}
\end{equation}
Here, $\theta$ denotes the filament orientation. 
Unlike the optimal filter in Equation~\ref{filter_HH2024}, the weights defined in Equations~\ref{real_space_filter} and~\ref{inverse_variance_weight} do not include correlations in the LSS noise covariance. The effect of these correlations is accounted for in the total variance of $\Gamma$, as described in Equation~\ref{eqn:total_err}.
The base variance $\sigma_{i,\rm base}^2$ is defined per shear component for the $i$th galaxy and consists of the intrinsic shape noise ($\sigma_{\rm shape}=0.25$ in this study), the measurement error ($\sigma_{i, \rm meas}$), and the halo-subtraction uncertainty ($\sigma_{i, \rm halo}$; see \textsection\ref{halo_subtraction}). We treat these noise terms as uncorrelated between galaxies. 
In this study, $\sigma_{i, \rm halo}$ is only included for the halo-subtracted analysis. For the non-subtracted orientation scan, we set $\sigma_{i, \rm halo}$=0.
For the variance of $\Gamma$, in addition to the base noise, we account for the correlated LSS contribution separately. The total variance of $\Gamma$ is defined as
\begin{equation}
\sigma^2_{\Gamma, \rm total} (\theta) = \sigma^2_{\Gamma, \rm base}(\theta) + \sigma^2_{\Gamma, \rm LSS}(\theta),
\label{eqn:total_err}
\end{equation}
where
\begin{equation}
\sigma^2_{\Gamma, \rm base} (\theta) = \frac{1}{W(\theta)^2}\sum_i \sigma_{i, \rm base}^2 \, \Psi_{i, \rm R}(\theta)^2,
\label{variance_base_noise}
\end{equation}
and $\sigma_{\Gamma, \rm LSS}(\theta)$ is estimated from mock LSS realizations (see \textsection\ref{LSS_mock_noise}).\footnote{For simplicity, we include the LSS contribution in the uncertainty of $\Gamma$, not in $w_i$. This choice leaves $\langle\Gamma\rangle$ unchanged in expectation and yields a more conservative variance estimate.}

\subsubsection{Mitigation of Host Halo Signal}\label{halo_subtraction}

\begin{figure}
\centering
\includegraphics[width = \columnwidth]{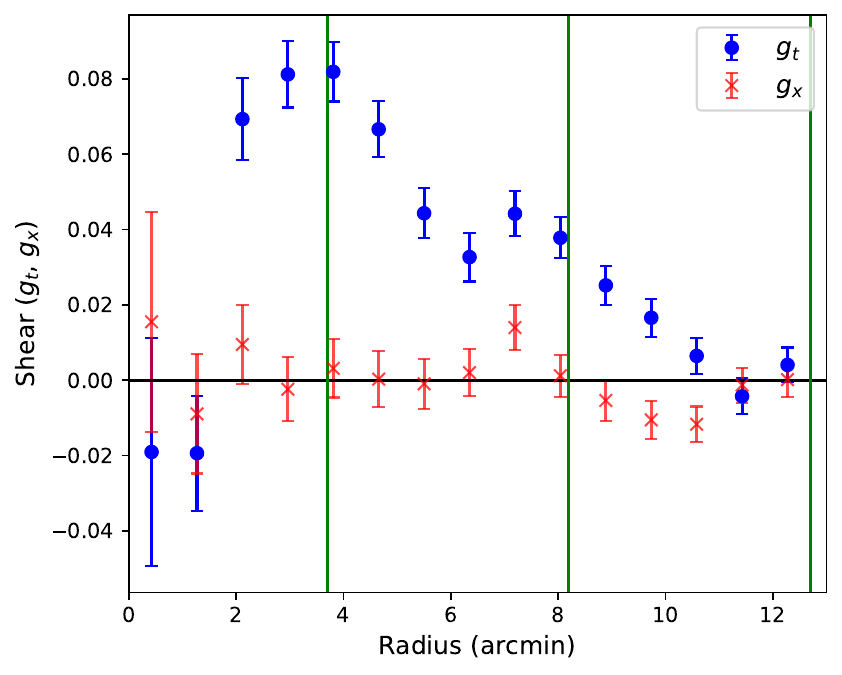}
\caption{Radially binned reduced shear profile to assess the halo contribution. Blue circles (red crosses) show the tangential component (cross component) of the reduced shear. Error bars indicate the standard uncertainties. Green vertical lines mark the inner and outer radial ranges adopted in the filament analysis.}
\label{fig:radial_gt}
\end{figure}

To mitigate the impact of the host cluster halo on filament property measurements, we subtract the halo signal from the observed shear field.
We first estimate the host contribution using the azimuthally averaged tangential shear profile $g_t$ (Figure~\ref{fig:radial_gt}). For each source galaxy, the corresponding radial $g_t$ value is projected onto the two shear components and subtracted from the observed ellipticities.
We quantify the uncertainty of this subtraction using the standard error of the binned profile, propagating it into the variance term $\sigma_{i,{\rm halo}}$ as detailed in \textsection\ref{matched_filter_stat}.
While this empirical scheme assumes azimuthal symmetry and leads to a degree of over-subtraction since $g_t$ inevitably includes contributions from the filaments themselves, it provides a first-order correction for the dominant host halo signal without requiring parametric model assumptions.

\subsubsection{Large-Scale Structure Noise Modeling}\label{LSS_mock_noise}
To estimate the LSS contribution to the measured shears,
we generate 1000 mock LSS convergence maps, $\kappa_{\rm LSS}$, as Gaussian random field (GRF) realizations derived from the angular convergence power spectrum $C_\kappa(\ell)$.\footnote{The GRF realizations capture the dominant LSS variance implied by $C_\kappa(\ell)$.  Any non-Gaussianity is expected to have a subdominant impact on our conclusions.} We compute $C_\kappa(\ell)$ using {\tt pyccl} \citep[][]{2019ApJS..242....2C}, and for simplicity, adopt a single source plane at $z=1$, the effective source redshift of our WL catalog (see \textsection\ref{background_source_redshift}).
For each realization, we convert $\kappa_{\rm LSS}$ into the corresponding shear field and evaluate the shear at the positions of the observed source galaxies to preserve the discrete and irregular galaxy sampling. We repeat the matched-filter measurement using the LSS-only shear field and define the LSS variance as the sample variance of $\Gamma_{\rm LSS}(\theta)$ over the 1000 realizations, which gives $\sigma^2_{\Gamma, \rm LSS}(\theta)$ in Equation~\ref{eqn:total_err}.

\subsection{Two-dimensional Filament Structure Detection: Stepwise 2D Tracing}\label{method:2d_filament}
The matched-filter method used in previous studies can only search directions that originate from a predefined reference point and therefore the inferred directions can vary with the adopted reference point.
Although this approach is efficient for identifying filament directions, several aspects of real filament geometry are hard to represent. For example, filamentary structures are not always well described as directions that extend radially from a single point. Local curvature is also difficult to follow with this framework. In this sense, the matched-filter method is effective for identifying direction candidates and providing a 1D statistic, but it is not designed to trace the 2D morphology of filaments.

To address these limitations, we introduce a method to reconstruct the 2D filament morphology by tracing filament directions locally. The main idea is to infer a filament path as a sequence of locally preferred directions, rather than summarizing it with a single direction from the matched-filter scan. 
Because the path is traced locally, the full structure is not required to pass through a single predefined point, making the method less sensitive to the reference point.
At each location, we scan orientations within a local window and move along the direction that yields the maximum S/N. We adopt a top-hat filter (i.e., a uniform kernel) to measure the signal, which reduces model dependence that can arise when adopting a specific filament profile. Iterating this approach produces a stepwise 2D trajectory of the filament in the shear field.
We apply the stepwise 2D tracing to the halo-subtracted shear field, since the cluster halo can contribute non-negligibly to the local alignment signal and bias the inferred directions toward radial patterns.

We implement the stepwise 2D tracing as follows. First, we initialize stream seeds on the boundary at $R=1~\mathrm{Mpc}$, and place the seeds at a uniform angular spacing of $1^\circ$. Starting from each seed, we measure the local S/N using the same matched-filter weighting scheme described in \textsection\ref{matched_filter_stat}, but with a top-hat kernel.
Although the analytic profile in Equation~\ref{filament_profile} could be used as the weighting kernel, its extended tails would assign weights to galaxies outside the local region of interest unless an additional truncation were introduced. We therefore adopt a top-hat kernel to keep the measurement local.
At every step, we scan orientations within a local aperture centered on the current point and choose the direction that maximizes the S/N of $\Gamma_{+}$. 
To ensure physical plausibility, we impose directional constraints: at the initial seeding step, the scan is restricted to be radially outward from the cluster center. For subsequent steps, we constrain the search direction to lie within $\pm45^\circ$ relative to the previous step to prevent unphysical sharp turns. This strategy aligns with cosmological simulations, indicating that filaments around clusters tend to extend radially outward \citep[e.g.,][]{2021MNRAS.502..714R}.

In this work, we adopt a rectangular aperture with a width of $0.8~\mathrm{Mpc}$, a length of $1.6~\mathrm{Mpc}$, a step size of $0.4~\mathrm{Mpc}$, and an angular sampling of $\Delta\theta=1^\circ$.
The termination criteria are defined as ${\rm S/N}(\Gamma_{+})>2$ and consistency of $\Gamma_{\times}$ with zero within $1\sigma$ uncertainties. We demonstrate the robustness of our results against variations in these geometric constraints and parameter choices in Appendix~\ref{robustness_step_by_step}.

To efficiently incorporate the LSS contribution, we employ a two-level strategy. We first generate stream candidates using the base noise $\sigma_{\Gamma,\rm base}$ (Equation~\ref{variance_base_noise}), and then re-evaluate the selected streams by including the LSS noise $\sigma_{\Gamma,\rm LSS}$ in the S/N computation. This approach avoids the computational cost of re-estimating $\sigma_{\Gamma,\rm LSS}$ at every step during the initial tracing.

\section{Results}\label{section:result}
In this section, we present the filament search and characterization results in three steps. In \textsection\ref{detect_filament_orientations}, we run the matched-filter orientation scan on the non-subtracted data. The scan is repeated on the halo-subtracted data to examine how the peaks change after halo subtraction. In \textsection\ref{filament_property_fitting}, we characterize the filament property 
using the halo-subtracted shear data. In \textsection\ref{2d_fil_dis}, we apply the stepwise 2D 
tracing analysis to the halo-subtracted data.

\subsection{Matched-filter Orientations}\label{detect_filament_orientations}
\begin{figure*}
\centering
\includegraphics[width=0.8\textwidth]{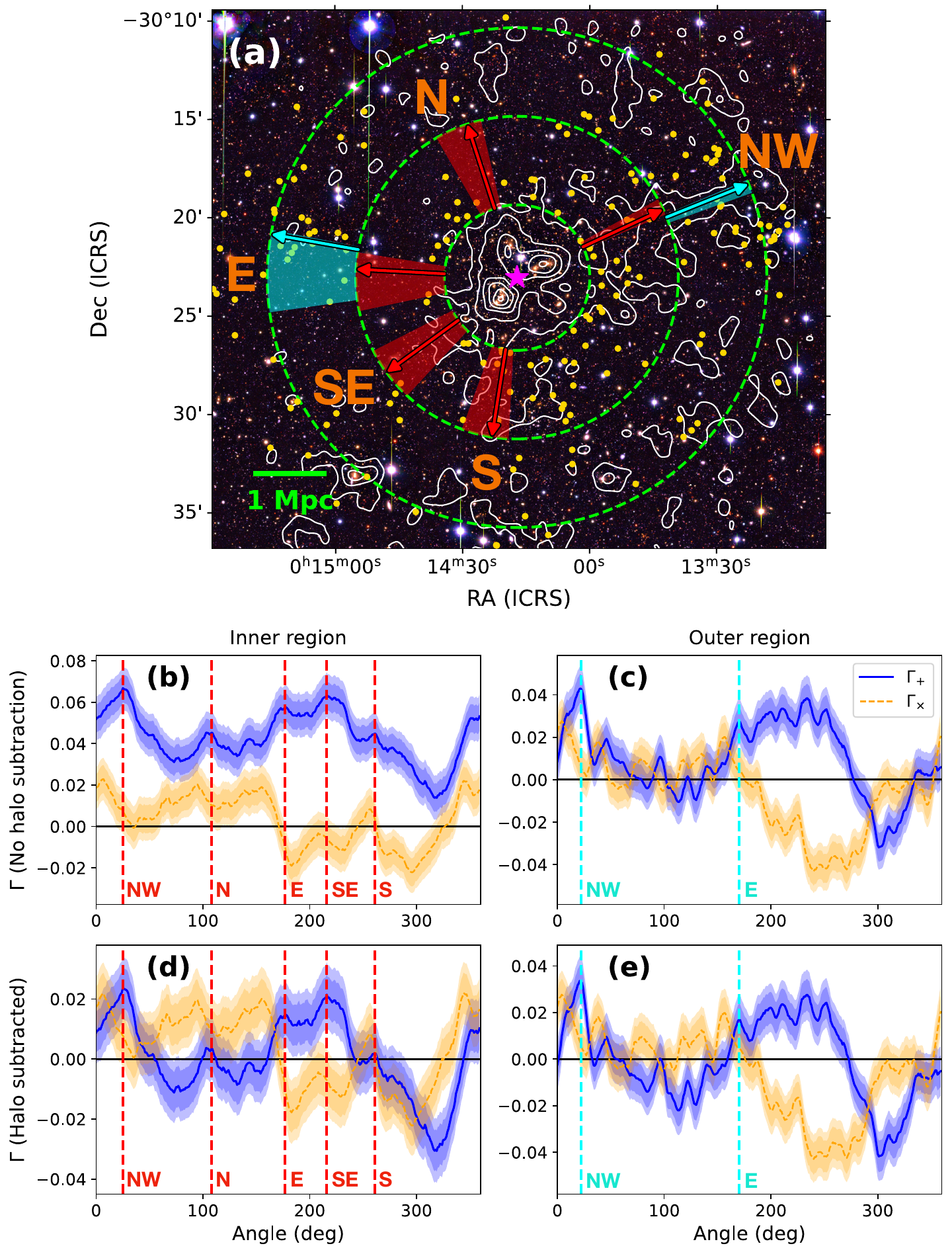} 
\caption{Filament detection using the matched-filter method. (a) Detected filaments around Abell 2744. The dashed green circles mark radii of 1, 2.2, and 3.4 Mpc, respectively. The red (cyan) arrows indicate the best-fit directions measured in the inner (outer) annulus. The labels NW, N, E, SE, and S denote the filament candidates discussed throughout the text and tables.
The shaded regions represent the 16th-84th percentile ranges from 1000 bootstrap realizations. The white contours show the reconstructed WL mass map at $\kappa=[0.12, 0.24, 0.36, 0.48, 0.6, 0.7]$. The yellow circles mark cluster member galaxies. The magenta star denotes the fiducial reference point adopted for the analysis. The colormap is the same as in Figure~\ref{fig:a2744_massmap_zoom}. The field of view is $\sim30\arcmin\times28\arcmin$. (b)-(e) Matched-filter statistics $\Gamma_+$ and $\Gamma_\times$ as a function of position angle. The blue solid (orange dashed) curves show the tangential (cross) components. The shaded regions indicate $1\sigma$ uncertainties without (dark) and with (light) LSS noise. The inclusion of LSS noise increases the uncertainty estimates by $\sim55\%$. The red (cyan) vertical dashed lines mark the detected filament directions in the inner (outer) annulus. Panels (b) and (c) are obtained from non-subtracted shears, whereas panels (d) and (e) are from halo-subtracted data.
} 
\label{fig:filament_direction}
\end{figure*}

In Figure~\ref{fig:filament_direction}, we present the filament direction candidates identified from the matched-filter method and the 1D statistics as a function of orientation. We adopt ${\rm S/N}>2$ for $\Gamma_{+}$ and require $\Gamma_{\times}$ to be consistent with zero within its $1\sigma$ uncertainty. 
The arrows in Figure~\ref{fig:filament_direction}(a) indicate the peak orientations that maximize the S/N measured from the 
non-subtracted WL catalog in the inner ($1~\mathrm{Mpc}<R<2.2~\mathrm{Mpc}$) and outer ($2.2~\mathrm{Mpc}<R<3.4~\mathrm{Mpc}$) annuli. We identify five candidates in the inner annulus (red arrows) and two in the outer annulus (cyan arrows). 
We refer to these filament candidates as NW, N, E, SE, and S according to their approximate directions.
Directional uncertainties are estimated from 1000 bootstrap realizations of the WL catalog and shown as the $16$–$84$ percentile ranges. Among the candidates, 
the NW and E filament candidates
are coherent between the two annuli. The inner and outer estimates for these candidates are consistent within the 1$\sigma$ (16th–84th percentile) ranges. 
The NW 
candidate has $\theta_{\rm inner}\sim25^\circ$ and $\theta_{\rm outer}\sim22^\circ$, while the 
E 
candidate has $\theta_{\rm inner}\sim177^\circ$ and $\theta_{\rm outer}\sim170^\circ$.

The scans are shown in Figures~\ref{fig:filament_direction}(b) and (c) for the inner and outer annuli, respectively, and in Figures~\ref{fig:filament_direction}(d) and (e) after subtracting the halo contribution. 
The inclusion of LSS noise increases the uncertainty estimates by $\sim 55\%$ compared to the case without LSS noise.
In all cases, the peak S/N decreases after halo subtraction, and the reduction is more prominent in the inner annulus. In particular, in the inner annulus, the peaks of the N and S candidates drop to ${\rm S/N}\lesssim0.5$ (see Table~\ref{tab:wide_table}). This indicates that a non-negligible fraction of the directional peaks can be driven by the host-halo shear, motivating our use of the halo-subtracted field for parameter inference in \textsection\ref{filament_property_fitting}.

Figure~\ref{fig:filament_direction}(a) also compares the candidate directions with the WL mass map and the distribution of spectroscopically confirmed cluster members. The NW 
candidate, which has the highest peak S/N and the smallest uncertainty in both annuli, follows the mass elongation seen in the WL mass map and is consistent with the cluster member galaxy distribution. The E
candidate aligns with the distribution of cluster members. Its bootstrap distribution is noticeably skewed. 
This suggests that the inferred E direction may be sensitive to the adopted reference point. We test this dependence in Appendix~\ref{offset_node_cluster} and compare the result with the stepwise-traced direction in \textsection\ref{2d_fil_dis}.
The remaining candidates show weaker or no obvious counterparts in these tracers, while a subset appears consistent with features suggested by the X-ray analysis. A detailed comparison with the X-ray morphology is presented in \textsection\ref{xray_analysis}.

\begin{table*}[!t]
    \centering
    \begin{threeparttable}
        \caption{Filament Properties in the Inner and Outer Annuli}
        \label{tab:wide_table}
        \begin{tabular*}{\textwidth}{@{\extracolsep{\fill}}cccccccc}
            \toprule
             & Filaments & Orientation$^{\rm b}$ & Peak S/N$^{\rm c}$ & $\kappa_0$ & $h_c$ & $\rho/\rho_b~^{\rm d}$ & Linear Mass Density$^{\rm e}$ \\
             & & (deg) & & & (Mpc) & & ($10^{13} \, \mathrm{M}_{\odot}~\mathrm{Mpc}^{-1}$) \\ 
            \midrule
            \multirow{6}{*}{Inner} & NW & $26^{+4}_{-3}$ & 6.2 (2.1) & $0.068^{+0.031}_{-0.028}$ & $0.205^{+0.170}_{-0.081}$ & $233^{+106}_{-95}$ & $2.57^{+3.51}_{-1.47}$\\
             & N & $108^{+12}_{-6}$ & 4.4 (0.5) & $0.015^{+0.018}_{-0.010}$ & $0.494^{+0.537}_{-0.369}$ & $53^{+62}_{-35}$ & $2.53^{+8.22}_{-2.30}$\\
             & E  & $177^{+13}_{-6}$ & 5.4 (1.5) & $0.032^{+0.012}_{-0.011}$ & $0.979^{+0.274}_{-0.413}$ & $111^{+41}_{-38}$ & $24.46^{+19.43}_{-15.73}$ \\
            & E (New)$^{a}$ & - & - & $0.035^{+0.025}_{-0.015}$ & $0.503^{+0.567}_{-0.309}$ & $122^{+87}_{-53}$ & $7.19^{+18.00}_{-5.54}$ \\
             & SE  & $217^{+12}_{-10}$ & 5.7 (1.9) & $0.062^{+0.044}_{-0.030}$ & $0.204^{+0.243}_{-0.119}$ & $215^{+151}_{-104}$ & $2.22^{+4.69}_{-1.66}$ \\
             & S  & $261^{+6}_{-10}$ & 4.3 (0.2) & $0.031^{+0.082}_{-0.025}$ & $0.095^{+0.494}_{-0.072}$ & $108^{+281}_{-88}$ & $0.23^{+1.92}_{-0.19}$ \\
            \cmidrule{1-8} 
            \multirow{3}{*}{Outer} & NW & $21^{+2}_{-3}$ & 4.1 (3.2) & $0.093^{+0.031}_{-0.028}$ & $0.237^{+0.156}_{-0.079}$ & $320^{+106}_{-98}$ & $4.76^{+5.24}_{-2.29}$ \\
             & E  & $171^{+17}_{-2}$ & 2.4 (1.5) & $0.099^{+0.082}_{-0.069}$ & $0.060^{+0.408}_{-0.032}$ & $343^{+282}_{-238}$ & $0.36^{+4.89}_{-0.25}$ \\
             & E (New)$^{a}$ & - & - & $0.032^{+0.031}_{-0.018}$ & $0.279^{+0.521}_{-0.175}$ & $109^{+106}_{-63}$ & $2.12^{+7.17}_{-1.71}$ \\
                                    
            \bottomrule
        \end{tabular*}
        \begin{tablenotes}
            \small 
            \item\textbf{Notes.}
            \item[a] ``New" indicates orientation derived from the stepwise tracing approach. See \textsection\ref{stepbystep_fitting} for details.
            \item[b] Median with 16th and 84th percentiles from 1000 bootstrap realizations.
            \item[c] Values in parentheses denote S/N after halo subtraction.
            \item[d] Density contrast relative to the mean matter density. Estimated assuming cylindrical symmetry within a radius of $4h_c$.
            \item[e] Integrated within a width of $2h_c$.
        \end{tablenotes}
    \end{threeparttable}
\end{table*}

\subsection{Properties of Detected Filaments}\label{filament_property_fitting}
\begin{figure*}
\centering
\includegraphics[width=\textwidth]{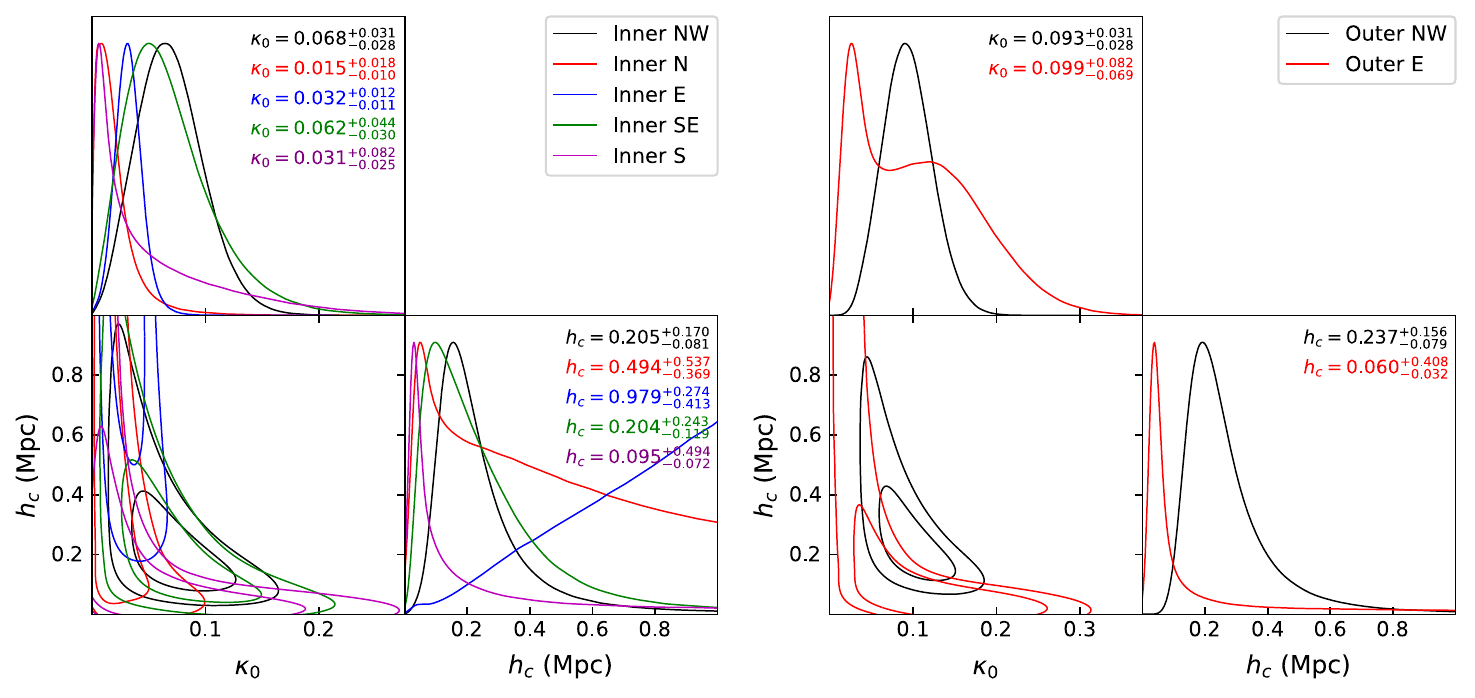} 
\caption{Posterior distributions of filament parameters from MCMC sampling. The left (right) panel shows the posterior distributions for the filament candidates in the inner (outer) annulus. The 1D curves are marginalized posteriors of $\kappa_0$ and $h_c$, and the 2D solid contours indicate the 68\% and 95\% posterior regions. The median and $1\sigma$ intervals are reported in each panel.} 
\label{fig:filament_posterior}
\end{figure*}

Figure~\ref{fig:filament_posterior} presents the posterior distributions for the filament candidates detected in \textsection\ref{detect_filament_orientations}. We sample the posteriors using MCMC with the analytic filament profile (Equation~\ref{filament_profile}) and use the halo-subtracted signal for parameter inference. In this section, we use a diagonal approximation to the LSS covariance in the likelihood (i.e., the LSS noise is assumed to be uncorrelated for the posterior sampling in this study).

Overall, most candidates in both annuli yield well-constrained posteriors, while some directions become prior-dominated or show strong degeneracy. 
For the E candidate, this occurs for $h_c$ in the inner annulus and for $\kappa_0$ in the outer annulus (see Table~\ref{tab:wide_table}).
We summarize the physical properties of the filament candidates inferred from our profile fitting in Table~\ref{tab:wide_table}. 

We compare the fitted filament parameters with previous simulation results. \citet{2014MNRAS.441..745H} applied the same analytic filament profile (Equation~\ref{filament_profile}) to simulated convergence maps and found that candidates classified as straight filamentary structures occupy a distinct region in the $(\kappa_0,\theta_c)$ plane (see their Figure 4). After converting our fitted $h_c$ values to angular scales at the redshift of Abell 2744, our median $(\kappa_0,\theta_c)$ values fall within the distribution of their straight-filament sample.
In addition, the inferred characteristic widths $h_c$ are consistent within uncertainties with the filament width reported in $z=0$ cosmological simulations \citep[$\sim 0.25~\mathrm{Mpc}$; ][]{2022A&A...661A.115G}. 
Regarding the density contrast, we estimate the filament density assuming cylindrical symmetry within a radius of $4h_c$. The NW candidate in both annuli and the SE candidate in the inner annulus show overdensities comparable to the values reported in the Coma cluster \citep{2024NatAs...8..377H}. The remaining orientations show relatively lower overdensities, consistent within $1\sigma$ with the estimates for cosmic filaments provided by \citet{2022A&A...661A.115G}\footnote{For the simulation comparison, we use the overdensity values recomputed and reported by \citet{2024NatAs...8..377H} from \citet{2022A&A...661A.115G}.}. Although the redshift difference ($z\sim0.3$ vs. $z=0$) limits strict quantitative comparison, these baselines provide a useful sanity check that our WL-inferred parameters lie in a physically plausible regime. 
Finally, the inferred linear mass densities are of the order of $\sim10^{13}~\mathrm{M}_{\odot}~\mathrm{Mpc}^{-1}$. These values are consistent in order of magnitude with the WL mass estimates of the filamentary structures in Abell 2744 reported by \citet{2015Natur.528..105E}, supporting the association of our detections with the previously reported large-scale features.

In \textsection\ref{detect_filament_orientations}, we find that the NW and E filament candidates
are coherent between the two annuli. For the NW candidate, 
the inferred $\kappa_0$ and $h_c$ are also consistent between the inner and outer annuli within $1\sigma$, supporting the consistency of this filament detection. In contrast, the E candidate
shows noticeable inconsistencies, including a skewed bootstrap distribution, an $h_c$ posterior pushed to the edge of the prior range in the inner annulus, and a bimodal $\kappa_0$ posterior in the outer annulus. 
The recentered scans in Appendix A show that inferred E orientation depends on the adopted reference point. We examine its impact on parameter inference further in \textsection\ref{2d_fil_dis} and \textsection\ref{stepbystep_fitting}.

\subsection{Stepwise Tracing of 2D Morphology}\label{2d_fil_dis}
\begin{figure*}
\centering
\includegraphics[width = \textwidth]{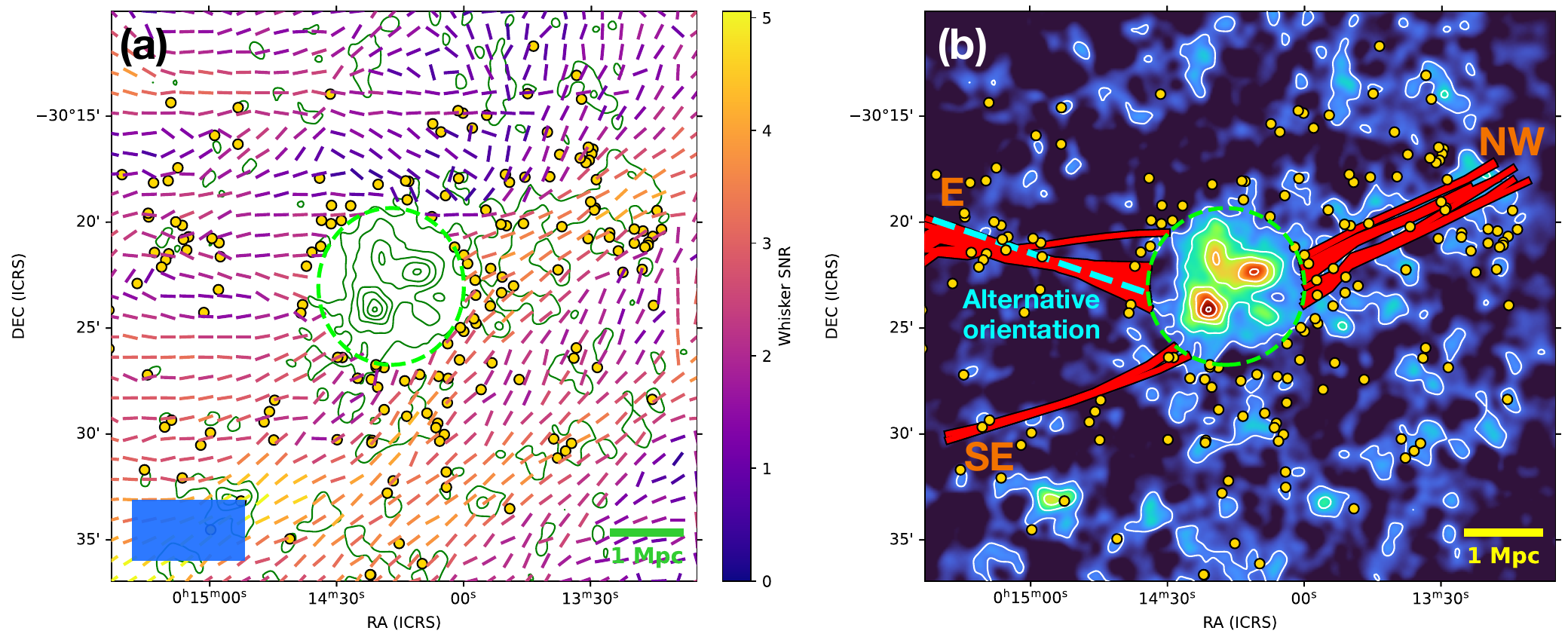}
\caption{Traced 2D filament structure around Abell 2744. (a) Whisker S/N plot for visualization. At each position, the whisker orientation corresponds to the angle that maximizes the local matched-filter S/N, with the color indicating the S/N. The blue rectangle in the lower left corner indicates the size of the top-hat filter used in the analysis. (b) Traced 2D filament paths (red curves) obtained from the stepwise tracing approach overlaid on the WL mass map shown as a color map. The dashed cyan line indicates the alternative orientation adopted for the alignment test in the filament fitting (see text). The labels NW, E, and SE denote the corresponding filament candidates. In both panels, the yellow circles mark cluster member galaxies, the lime circle denotes the inner-annulus boundary, and the contours show the reconstructed WL mass map, as in Figure~\ref{fig:filament_direction}.}
\label{fig:2d_filament}
\end{figure*}

In Figure~\ref{fig:2d_filament}, we present a ``whisker" visualization of the local shear alignment and the traced two-dimensional filament morphology from our stepwise tracing procedure. We use the halo-subtracted shear field to mitigate the host-halo contribution in the local alignment signal. In the left panel, each whisker orientation is set to the angle that maximizes the local matched-filter S/N at that location. The $\Gamma_{\times}$ consistency requirement is not applied for this whisker map. The whisker map visualizes the local directional preference in the shear field and is not used in the stepwise tracing method.

The whisker map reproduces trends that are broadly consistent with the matched-filter analysis. In particular, it shows a clear directional preference toward the northwestern side, in agreement with the matched-filter result that the NW candidate
is the most significant. The field also shows extended features toward the eastern and southeastern sides. The southwestern part of the field shows a diagonal pattern, with directions roughly aligned from southeast to northwest.

In Figure~\ref{fig:2d_filament}(b), we show the filament streams traced by the stepwise tracing algorithm described in \textsection\ref{method:2d_filament}. For stream selection, we adopt ${\rm S/N}(\Gamma_{+})>2$ and $\Gamma_{\times}$ to be consistent with zero within $1\sigma$ uncertainties at all steps along each stream. With this baseline criterion, we identify three prominent filament paths on the northwestern, eastern, and southeastern sides. The stepwise tracing method recovers paths associated with the NW and E filament candidates identified in both annuli by the fiducial matched-filter scan.\footnote{The streams have different detection strengths. The eastern stream falls below the detection threshold of ${\rm S/N}=3$, while other streams remain above this threshold.} We do not recover comparable stepwise-traced paths for the N and S candidates 
in the halo-subtracted field. This is consistent with their weak matched-filter signals after halo subtraction (Figure~\ref{fig:filament_direction}(d) and (e)).

The stepwise tracing method also recovers a path associated with the SE filament candidate. In the matched-filter analysis, the SE candidate is detected in the inner annulus, whereas the outer-annulus peak at a similar angle is excluded because $\Gamma_{\times}$ is not consistent with zero (see Figure~\ref{fig:filament_direction}). 
This suggests that the matched-filter filament approach may not be well-suited in this direction. In particular, the inner-annulus matched-filter direction of the SE candidate
points toward the region where \citet{2015Natur.528..105E} reported foreground structures. Our stepwise tracing method traces filamentary structures on the northern side of that region, and we discuss this comparison in \textsection\ref{xray_analysis} (see Figure~\ref{fig:combien_WL_xray}).

As discussed in \textsection\ref{filament_property_fitting} and Appendix~\ref{offset_node_cluster}, the fiducial matched-filter scan can yield a direction that does not fully represent the local geometry when the adopted reference point does not lie along the inward continuation of the structure.
Motivated by the traced morphology, we define an alternative elongation direction for the 
E filament candidate
and mark it as the dashed cyan line in Figure~\ref{fig:2d_filament}(b). Along this alternative direction, we find an overdensity in the WL mass map, which supports the interpretation that the best-fit direction from the fiducial matched-filter scan does not always overlap with the locally preferred filament orientation. This alternative direction is also closely aligned with the E direction recovered from the matched-filter scan centered on the southern mass peak in Appendix~\ref{offset_node_cluster}. 
The closer agreement in this case is naturally explained by filaments extending outward from massive halos.
We will discuss this alternative direction further in \textsection\ref{stepbystep_fitting}.

\section{Discussion}\label{section:discussion}

\subsection{Impact of the Reference Point on Matched-filter Fitting}\label{stepbystep_fitting}

Building on the reference-point test in Appendix~\ref{offset_node_cluster}, we quantify how the difference between the fiducial matched-filter and locally preferred E orientations affects the filament fit using two complementary criteria: the Bayesian evidence ($\ln Z$) and the stability of the inferred filament parameters.
We first compare the Bayesian evidence in \textsection\ref{eval_evidence}, followed by an analysis of the posterior stability in \textsection\ref{comp_post_filament}.

\subsubsection{Bayesian Evaluation of Filament Orientations}\label{eval_evidence}
\begin{figure}
\centering
\includegraphics[width = \columnwidth]{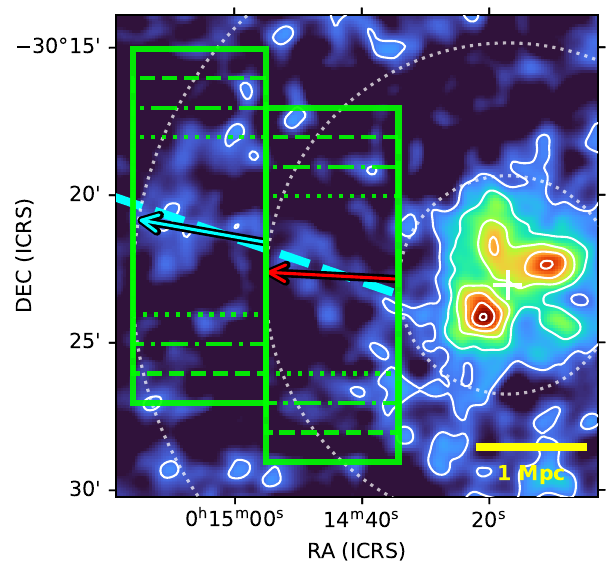}
\caption{Analysis windows for the Bayesian evidence calculation. The background mass map and contours are identical to those shown in Figure \ref{fig:2d_filament}. The lime rectangular boxes denote the analysis windows used for the evidence calculation, where different line styles represent varying window half-widths of $3\arcmin$, $4\arcmin$, $5\arcmin$, and $6\arcmin$. The arrows indicate the filamentary direction identified by the fiducial matched-filter scan, while the cyan dashed line represents the alternative orientation determined by the stepwise tracing method. In this zoomed view, the stepwise tracing orientation aligns more closely with the local mass overdensities compared to the fiducial matched-filter orientation.}
\label{fig:evidence_eval_window}
\end{figure}

\begin{deluxetable}{llcccc}[ht]
\tablecaption{Log Bayes factor ($\Delta\ln Z$) for Filament Orientation Models \label{tab:evidence_comparison}}
\tablewidth{0pt}

\tablehead{
    \colhead{Region} & \multicolumn{4}{c}{Box Width$^a$} \\ 
    \cline{2-5}
    \colhead{} & \colhead{$\pm 3\arcmin$} & \colhead{$\pm 4\arcmin$} & \colhead{$\pm 5\arcmin$} & \colhead{$\pm 6\arcmin$}
}

\startdata
Inner & +24.4 & +29.2 & +79.9 & +122.9 \\ 
\hline
Outer & +51.6 & +57.8 & +49.2 & +73.9 \\ 
\enddata

\tablecomments{Log Bayes factor is calculated as $\Delta \ln Z = \ln Z_{\text{New}} - \ln Z_{\text{Old}}$, where $\ln Z_{\text{Old}}$ corresponds to the fiducial matched-filter orientation, and $\ln Z_{\text{New}}$ to the stream-tracing orientation.
$^a$The transverse extent described in Figure~\ref{fig:evidence_eval_window}.}
\end{deluxetable}

In \textsection\ref{filament_property_fitting}, we observed that the E filament candidate 
yields unstable parameter constraints, despite the fiducial matched-filter analysis identifying a coherent orientation in both the inner and outer annuli. 
The recentered tests show that a closely aligned E direction can be recovered by changing the matched-filter reference point. The discrepancy reflects the limitation that the global filter axis must pass through the adopted reference point.

Motivated by the traced morphology described in \textsection\ref{2d_fil_dis}, we perform an alternative alignment test using Bayesian evidence computed with {\tt PyMultiNest} \citep{2014A&A...564A.125B}. To ensure a statistically valid comparison on the same dataset, we enforce an identical galaxy sample for both the matched-filter and stepwise tracing orientations by restricting the analysis to galaxies within fixed rectangular windows (see Figure~\ref{fig:evidence_eval_window}). 
The analytic filament profile used here is constant along the filament axis and varies only with the perpendicular distance from the axis (see Equation~\ref{filament_profile}). Since sector or wedge-shaped regions would sample an increasingly broad transverse region as the distance from the reference point increases, we use fixed rectangular windows to avoid this widening.
These windows are configured to cover the E filament candidate, with the inner window centered at $y=0$ and the outer window shifted to $y=+2\arcmin$ relative to the reference point. We vary the window width to verify the robustness of our results against the specific window choice. In all tested cases, the locally preferred orientation inferred from our stepwise tracing method is favored over the fiducial matched-filter orientation, 
yielding $\Delta\ln Z \equiv \ln Z_{\rm New}-\ln Z_{\rm Old} \ge 24$ (Table~\ref{tab:evidence_comparison}). This corresponds to \emph{very strong} support according to the \citet{kass1995bayes} criterion.

\subsubsection{Comparison of Posterior Distributions}\label{comp_post_filament}
\begin{figure*}
\centering
\includegraphics[width=\textwidth]{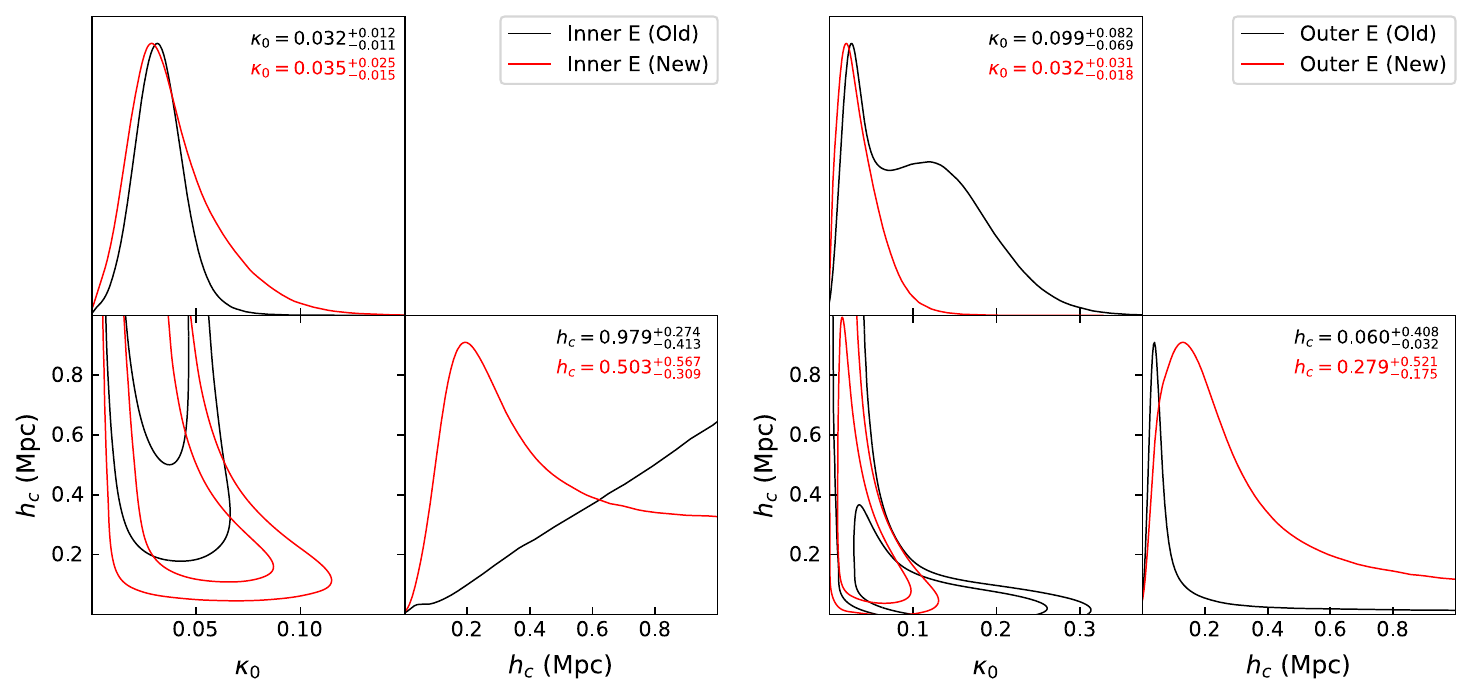} 
\caption{Posterior comparison between the fiducial matched-filter scan and the stepwise tracing method orientations for the E filament candidate. The 1D curves and 2D solid contours indicate the marginalized posteriors and 68\% and 95\% posterior regions, respectively. The median and $1\sigma$ intervals are reported in each panel. Here, ``Old" denotes the orientation that maximizes the fiducial radial matched-filter scan, whereas ``New" denotes the elongation direction inferred from the stepwise tracing method.} 
\label{fig:filament_posterior_compare}
\end{figure*}

Following the validation of the directions obtained by our stepwise tracing method, we perform a re-fitting of the E filament candidate along the elongation direction inferred from the stream tracing (cyan dashed line in Figure~\ref{fig:2d_filament}(b)). We repeat the MCMC sampling using the same analytic filament profile and data selection as described in \textsection\ref{filament_property_fitting}, but with the filter axis aligned to this new direction. Figure~\ref{fig:filament_posterior_compare} compares the posterior distributions for the E filament candidate obtained using the fiducial matched-filter scan orientation (``Old") and the stream-tracing direction (``New") for both the inner and outer annuli.

The stream-based orientation yields posteriors that are better constrained than those from the fiducial matched-filter orientation. Especially, the constraint on $h_c$ becomes less prior-dominated in the inner annulus, and the posterior of $\kappa_0$ no longer shows the strong bimodality seen in the fiducial case in the outer annulus. In addition, the fitted parameters become more consistent between the two annuli. With the stream-based orientation, the inner and outer posteriors of both $\kappa_0$ and $h_c$ agree within $1\sigma$, whereas the fiducial matched-filter orientation shows discrepancies between the annuli.

These results show that an orientation mismatch associated with the reference point choice can lead to unstable or inconsistent parameter inference. The stepwise tracing method recovers a locally coherent direction and therefore provides a characterization that is less sensitive to the choice of a single reference point.
Figure~\ref{fig:evidence_eval_window} also provides a qualitative visual inspection of this complementarity. While the fiducial matched-filter orientation (arrows) captures the global overdensity trend, the stepwise-tracing direction more closely follows the local overdensity ridges in the WL mass map. 
The improved agreement after recentering on the southern mass peak also reflects the tendency of filaments to extend outward from massive halos.

\subsection{Comparison with X-ray Analysis}\label{xray_analysis}
\begin{figure*}
\centering
\includegraphics[width=\textwidth]{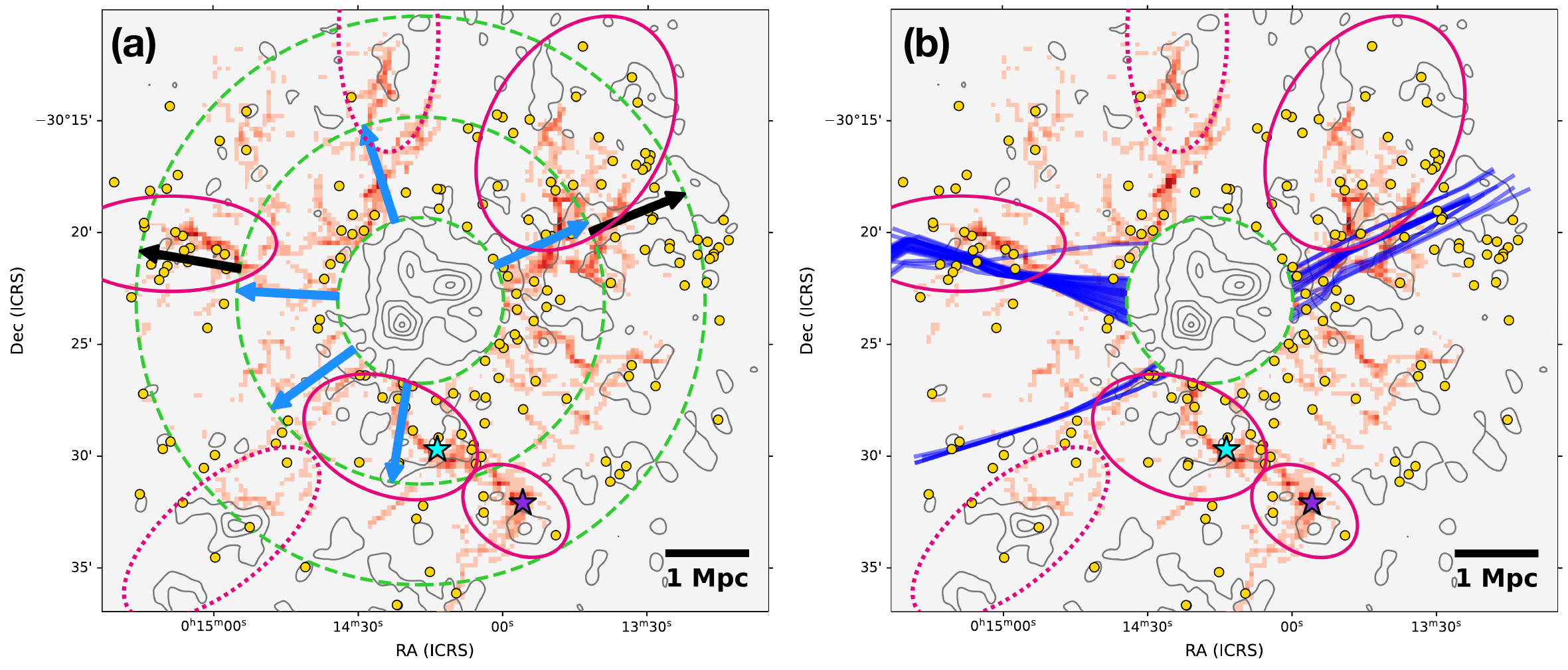} 
\caption{Filamentary structure comparison between WL and X-ray analyses. (a) Comparison to the best-fit filament orientations from the fiducial matched-filter scan. The arrows and dashed circles are the same as in Figure~\ref{fig:filament_direction}. (b) Comparison to the traced 2D filaments. The blue streams and dashed circle are the same as in Figure~\ref{fig:2d_filament}.
In both panels, the yellow circles indicate the cluster member galaxies. The solid (dotted) ellipses mark the identified filamentary (foreground and background) structures from X-ray analysis in \citet{2015Natur.528..105E}. The red intensity shows regions with filament probability $p>0.1$ from the T-REx algorithm in \citet{2024A&A...692A.200G}. The stars represent the cluster candidates from the WaZP cluster catalog. The contours show the reconstructed WL mass map, as in Figure~\ref{fig:filament_direction}.}
\label{fig:combien_WL_xray}
\end{figure*}

Previous XMM-Newton studies of Abell 2744 have reported diffuse large-scale structures around the cluster \citep{2015Natur.528..105E, 2024A&A...692A.200G}. In this section, we compare the filamentary structures from our WL analysis with X-ray results in previous studies. Figure~\ref{fig:combien_WL_xray} overlays our WL results with two X-ray tracers based on XMM-Newton data. The ellipses indicate diffuse structures reported by \citet{2015Natur.528..105E}. The solid ellipses were included in their analysis, while the dotted ellipses were excluded. The red intensity shows the filament probability map from \citet{2024A&A...692A.200G}. They used the Tree-based Ridge Extractor \citep[T-REx;][]{2020A&A...637A..18B} algorithm to map the probability. We show regions with filament probability $>0.1$, and the intensity is renormalized for visualization. The arrows in Figure~\ref{fig:combien_WL_xray}(a) indicate the best-fit filament orientations from our matched-filter scan, and the blue streams in Figure~\ref{fig:combien_WL_xray}(b) show the 2D filament morphology traced by our stepwise tracing approach.

The diffuse X-ray structures reported by \citet{2015Natur.528..105E} agree with our WL-based filament candidates in most cases, with a few differences. The NW and E candidates have filamentary counterparts in both analyses, although the NW feature shows a small positional offset between the WL and X-ray tracers. Our matched-filter scan also yields a signal for the N and S candidates. After halo subtraction, however, these features become weak (S/N$\lesssim0.5$). \citet{2015Natur.528..105E} excluded the north due to a background galaxy concentration. Our scan also yields a signal for the SE candidate toward a structure excluded from their analysis due to its large velocity offset from the cluster core. We do not obtain a clear southwest detection in our matched-filter scan, despite the diffuse feature reported in their X-ray analysis. In the stream-tracing map, the path associated with the E candidate traces the X-ray structures, while the paths associated with the NW and SE candidates show spatial offsets relative to the reported diffuse regions.

We also examine the T-REx filament probability map from \citet{2024A&A...692A.200G}. The high-probability ridges generally align with the directions identified by our matched-filter scan, whereas the agreement is weaker toward the south to southwest. The stream-tracing map shows the same trend. It follows the probability ridges in the eastern and northeastern regions, but we do not identify a robust southern stream feature. 
The T-REx comparison favors the eastern/northeastern connectivity indicated by the WL-based directions, although the southern side is not robustly supported by our WL results.

Both the diffuse structures reported by \citet{2015Natur.528..105E} and the T-REx probability map show a south-southwest feature, while our WL-based detection in this direction is weak. 
In the same area, we find two line-of-sight systems from the DES-Y6 WaZP catalog\footnote{\url{https://des.ncsa.illinois.edu/releases/y6a2/Y6cluster-wazp}} \citep{2025arXiv250705360B} with spectroscopic redshifts (marked by stars in Figure~\ref{fig:combien_WL_xray}; $z_{\rm spec}=0.1056$ and 0.4964)\footnote{The purple star corresponds to WAZP DES Y6 J001356.0$-$303205.6 ($z_{\rm spec}=0.1056$; RA$=3.48323003^\circ$, Dec$=-30.53489907^\circ$). The cyan star corresponds to WAZP DES Y6 J001413.7$-$302943.0 ($z_{\rm spec}=0.4964$; RA$=3.55698072^\circ$, Dec$=-30.49526570^\circ$).}.
These two cluster candidates lie along the south–southwest feature, and projection effects could contribute to the weaker WL signal in that direction.
However, since the WaZP catalog is identified in photometric-redshift space, a dedicated follow-up would be needed to assess whether these systems are responsible for the south-southwest feature.

Overall, the agreement between WL-based filament directions and X-ray diffuse structures supports that a substantial fraction of the detected filamentary signals trace large-scale structures connected to Abell 2744. The remaining discrepancies in a few directions suggest that a direct one-to-one comparison can be limited for this cluster, since WL and X-ray measurements probe different components and the complex merger state of the system. This cross-check therefore suggests the need for further multi-wavelength constraints, such as deeper X-ray observations and spectroscopy, to investigate the discrepant directions and to strengthen the physical interpretation of the filament candidates.

\subsection{Surrounding Cluster Candidates}\label{surrounding_cluster}
\begin{figure}
\centering
\includegraphics[width = \columnwidth]{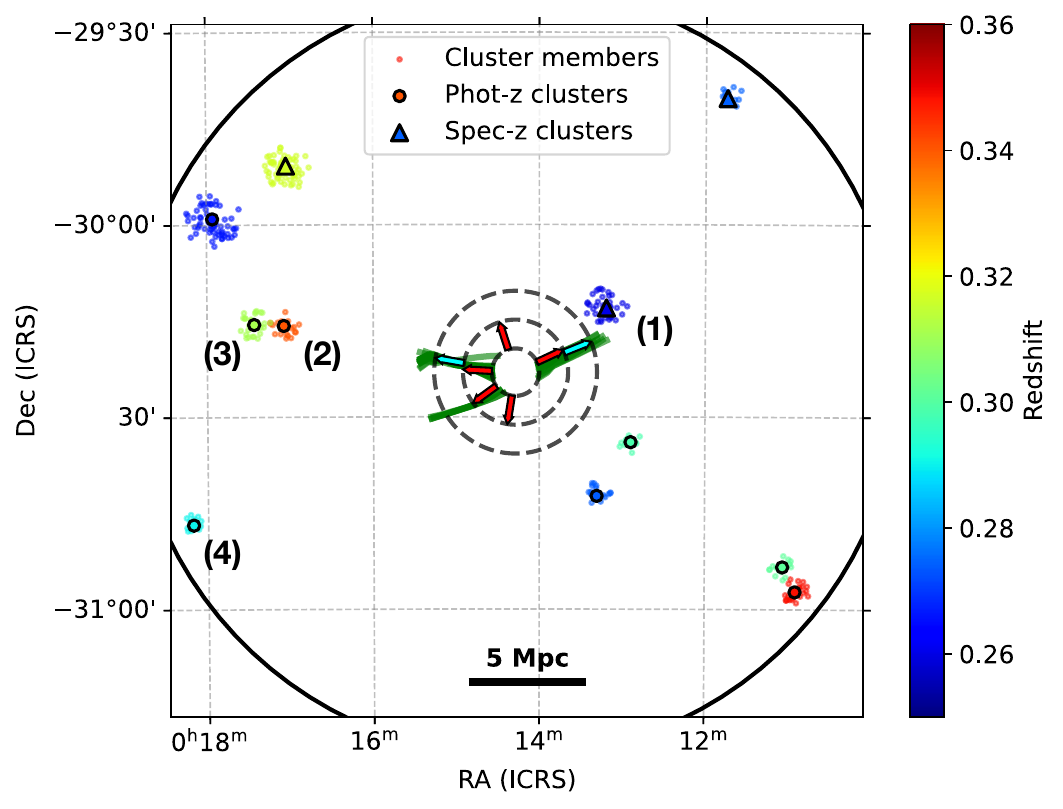}
\caption{Surrounding cluster candidates from the WaZP galaxy cluster catalog of the Dark Energy Survey. We select the cluster candidates within 1 deg of the fiducial reference point and within $\Delta z=\pm0.05$ of the redshift of Abell 2744. Small points show cluster member galaxies from the catalog. Filled circles mark WaZP cluster candidates with photometric redshifts, and triangles indicate systems with spectroscopic redshifts. We label the four candidates discussed in \textsection\ref{surrounding_cluster} as (1)–(4). The arrows, streams, and dashed circles are the same as in Figure~\ref{fig:filament_direction} and Figure~\ref{fig:2d_filament}.}
\label{fig:surrounding_clusters}
\end{figure}

In the large-scale structure of the Universe, matter forms a cosmic web in which dense nodes are connected by filaments \citep[e.g.,][]{1996Natur.380..603B, 2021MNRAS.502..714R}. If a filament orientation inferred from our analysis is physically meaningful, one may expect other nodes (i.e., galaxy clusters) to be found along a similar direction. To check this qualitatively, we use the WaZP galaxy cluster catalog from the Dark Energy Survey \citep{2025arXiv250705360B} used in \textsection\ref{xray_analysis}. 

In Figure~\ref{fig:surrounding_clusters}, we show WaZP cluster candidates within an angular separation between $12.7\arcmin$ and $1^\circ$ from our fiducial reference point, with richness $>10$ and redshifts within $\Delta z=\pm0.05$ of the cluster redshift ($z_{\rm cl}=0.308$). As an environmental cross-check, we use the WaZP catalog to see whether surrounding cluster candidates align with our inferred directions.
We find four cluster candidates along the directions of the NW, E, and SE filament candidates: 
(1) $\sim0.3^\circ$ (northwest; $z_{\rm spec}=0.2601$), (2) and (3) $\sim0.6^\circ$ (east; $z_{\rm phot}=0.3399$ and 0.3109), and (4) $\sim0.9^\circ$ (southeast; $z_{\rm phot}=0.2900$). The inferred filamentary structure directions are broadly consistent with the distribution of nearby cluster candidates.

\subsection{Alignment with the Merger Scenario}
Recently, \citet{2024ApJ...961..186C} reconstructed the lensing mass distribution of Abell 2744 by combining JWST SL and WL. They reported diffuse, elongated mass features (``mass bridges") and suggested that these structures may trace merger axes, consistent with the orientation of the radio relics. The NW, E, and SE filament candidates 
identified in our filament analysis show a qualitative alignment with the merger axes discussed in \citet[][Figure 14 therein]{2024ApJ...961..186C}. This alignment is seen both in the matched-filter scan after halo subtraction and in the stream tracing.

The northwest-southeast filament direction is consistent with the orientation of the mass bridge connecting the southern and northwest mass peaks in \citet{2024ApJ...961..186C}. The E filament candidate 
is not exactly aligned with the mass bridge between the northern and northwest mass peaks reported by \citet{2024ApJ...961..186C}, but it shares a broadly consistent orientation. 
This overall consistency suggests that the detected filament directions may trace the same large-scale connectivity or anisotropic accretion direction, rather than being random alignments. 

Meanwhile, the south and southwest directions suggested by the X-ray morphology are not comparably prominent in our halo-subtracted WL analysis. This mismatch may reflect tracer differences and/or line-of-sight projection. Given the complex merger state of Abell 2744, we treat the alignment between our WL filament directions and the merger axes as a qualitative hint that at least some of the filament candidates may trace the merger geometry. The inferred directions may therefore provide a complementary geometric indicator for interpreting the merger configuration.

\section{Conclusion}\label{section:conclusion}
We present a WL analysis of filamentary structures around Abell 2744 using Magellan/MegaCam data, including the host-halo subtraction and its uncertainty. We identify filament direction candidates using a matched-filter technique and introduce a new stepwise tracing method to trace the two-dimensional filament morphology in this study.

The fiducial matched-filter scan shows five orientation candidates in the inner annulus and two in the outer annulus, with the NW and E filament candidates detected in both annuli.
We fit the filament model using the fiducial matched-filter orientations of these candidates. After halo subtraction, the matched-filter signals for the N and S filament candidates 
drop to ${\rm S/N}\lesssim0.5$, while the NW, E, and SE filament candidates 
retain non-negligible signals. The NW candidate 
yields consistent constraints across the two annuli, whereas the E candidate 
exhibits less stable fits and a larger inner-outer discrepancy. 
Additional matched-filter scans centered on each of the three lensing mass peaks show that the inferred directions depend on the adopted reference point. The scan centered on the southern mass peak broadly recovers the NW and E directions, with its E direction closely aligned with the stepwise-traced path. This improved agreement further shows that filaments extend from massive halos as physically expected.

Using the stepwise tracing method, we reconstruct the 2D filament morphology and identify three dominant paths associated with the NW, E, and SE filament candidates. 
The method traces locally coherent directions without requiring the full structure to pass through a single predefined point and is therefore less sensitive to the reference point choice. The SE path illustrates this advantage because its inward continuation does not pass through any of the three mass peaks and it is not recovered in the recentered scans.
For the E candidate, the 2D morphology deviates from the fiducial matched-filter best-fit angle. When we re-fit the filament properties using the stepwise traced direction, the posterior constraints become more stable, the goodness-of-fit improves, and the inner and outer results become more consistent. 
This demonstrates that the stepwise tracing method can provide complementary information to the matched-filter scan, particularly when the filament geometry is not well captured by a single global orientation.

To cross-check our WL results, we compare them with X-ray tracers from previous XMM-Newton analyses. The overall correspondence supports that a substantial fraction of the detected signals trace large-scale structures connected to Abell 2744, while discrepancies in a few directions highlight the limitations of direct one-to-one comparisons between total mass and hot-gas tracers in a complex merging system. 
As an additional check in the outskirts, we examine the WaZP cluster candidates and find several plausible systems aligned with our filament directions. Our results are also qualitatively consistent with the merger-axis geometry proposed by \citet{2024ApJ...961..186C}.

Upcoming wide-area WL surveys, such as Euclid, the Rubin Observatory LSST, and Roman, will provide wide-field WL data for large cluster samples, enabling statistical studies of filament candidates across environments and out to larger radii. Our stepwise tracing method can complement the matched-filter scans by following filaments in two dimensions while allowing for locally varying directions. With complementary data such as X-ray observations and spectroscopy, these efforts can be extended into a multi-wavelength framework to probe the nature of filaments.

\begin{acknowledgments}
We thank Takahiro Morishita for sharing the Magellan/MegaCam data used in this work and for helpful comments on the manuscript. We also thank Wonki Lee and Kim HyeongHan for helpful discussions.
This paper includes data gathered with the 6.5 meter Magellan Telescopes located at Las Campanas Observatory, Chile. 
M. J. Jee acknowledges support for the current research from the National Research Foundation (NRF) of Korea under the programs 2022R1A2C1003130. SC acknowledges this research was supported by Basic Science Research Program through the NRF funded by the Ministry of Education (No. RS-2024-00413036).
\end{acknowledgments}

\software{Astropy \citep{astropy2013, 2018AJ....156..123A, 2022ApJ...935..167A}, emcee \citep{2013PASP..125..306F}, Matplotlib \citep{matplotlib2007}, NumPy \citep{harris2020array}, pyccl \citep{2019ApJS..242....2C}, PyMultiNest \citep{2014A&A...564A.125B}, PyTorch \citep{NEURIPS2019_bdbca288}, SciPy \citep{scipy2020}, SCAMP \citep{bertinscamp}, SWARP \citep{bertinswarp}, SExtractor \citep{1996A&AS..117..393B}}

\appendix
\section{Re-centering the Matched-Filter Scan}\label{offset_node_cluster}
\begin{figure*}
\centering
\includegraphics[width = 0.9\textwidth]{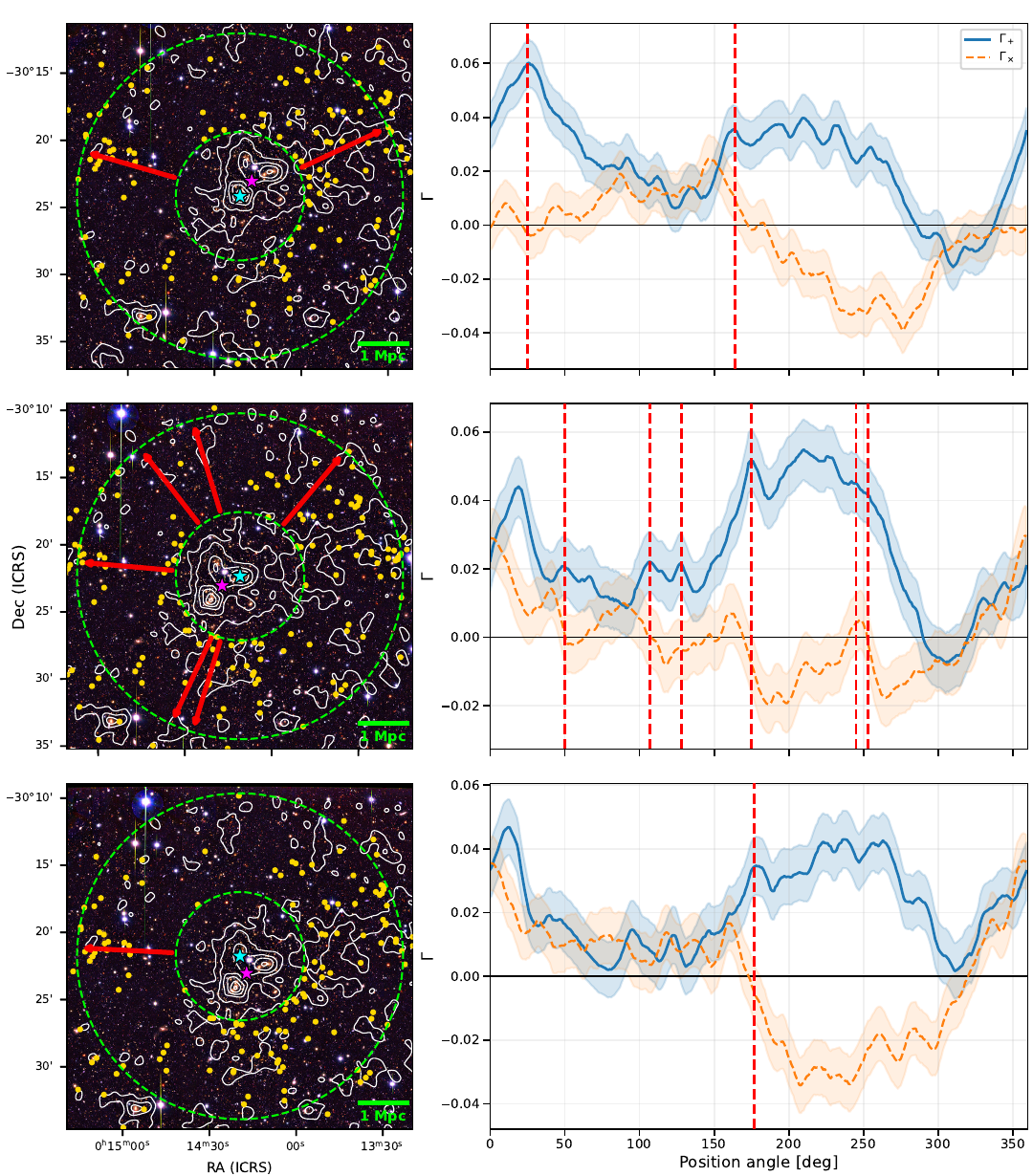}
\caption{Matched-filter scans centered on the southern (top), northern (middle), and northwestern (bottom) lensing mass peaks. The left panels show the detected directions over the WL mass map. The cyan star marks the reference point used in each scan, while the magenta star marks the fiducial geometric center. The dashed green circles indicate radii of 1.3 and 3.3 Mpc, respectively. The yellow circles mark cluster member galaxies. The red arrows indicate the best-fit directions. The right panels show the matched-filter statistics $\Gamma_{+}$ and $\Gamma_{\times}$ as a function of position angle obtained from non-subtracted shears, as in Figure~\ref{fig:filament_direction}.}
\label{fig:filament_direction_offset}
\end{figure*}

To test the dependence of the inferred directions on the adopted reference point, we repeated the matched-filter analysis using each of the three lensing mass peaks as the center on the non-subtracted shear field. For each scan, we applied the same detection criteria as in the fiducial analysis. Figure~\ref{fig:filament_direction_offset} presents the resulting directions and matched-filter statistics. Because the reference point changes, the original annular boundaries ($1.0-2.2$ and $2.2-3.4 ~\rm{Mpc}$) no longer map onto the same usable area on the sky. We therefore adopt an adjusted radial range ($1.3-3.3 ~\rm Mpc$).

The inferred directions vary with the adopted reference point. Especially, the scan centered on the southern mass peak, which is the most massive of the three, broadly recovers the NW and E directions. The E direction from this scan is closely aligned with the locally preferred direction identified by the stepwise tracing method. This supports the interpretation that the difference between the fiducial matched-filter and stepwise E directions mainly reflects the choice of the reference point. 
The improved agreement for the southern mass peak further suggests that filaments extend from massive halos.
None of the three recentered scans recovers the stepwise-traced SE direction. Although this path appears nearly straight over the observed region, its inward continuation does not pass through any of the three mass peaks. These tests demonstrate that the global matched-filter directions inferred for Abell 2744 are sensitive to the adopted reference point.

\section{Stepwise Tracing Robustness Tests}\label{robustness_step_by_step}
\begin{figure*}
\centering
\includegraphics[width = 0.85\textwidth]{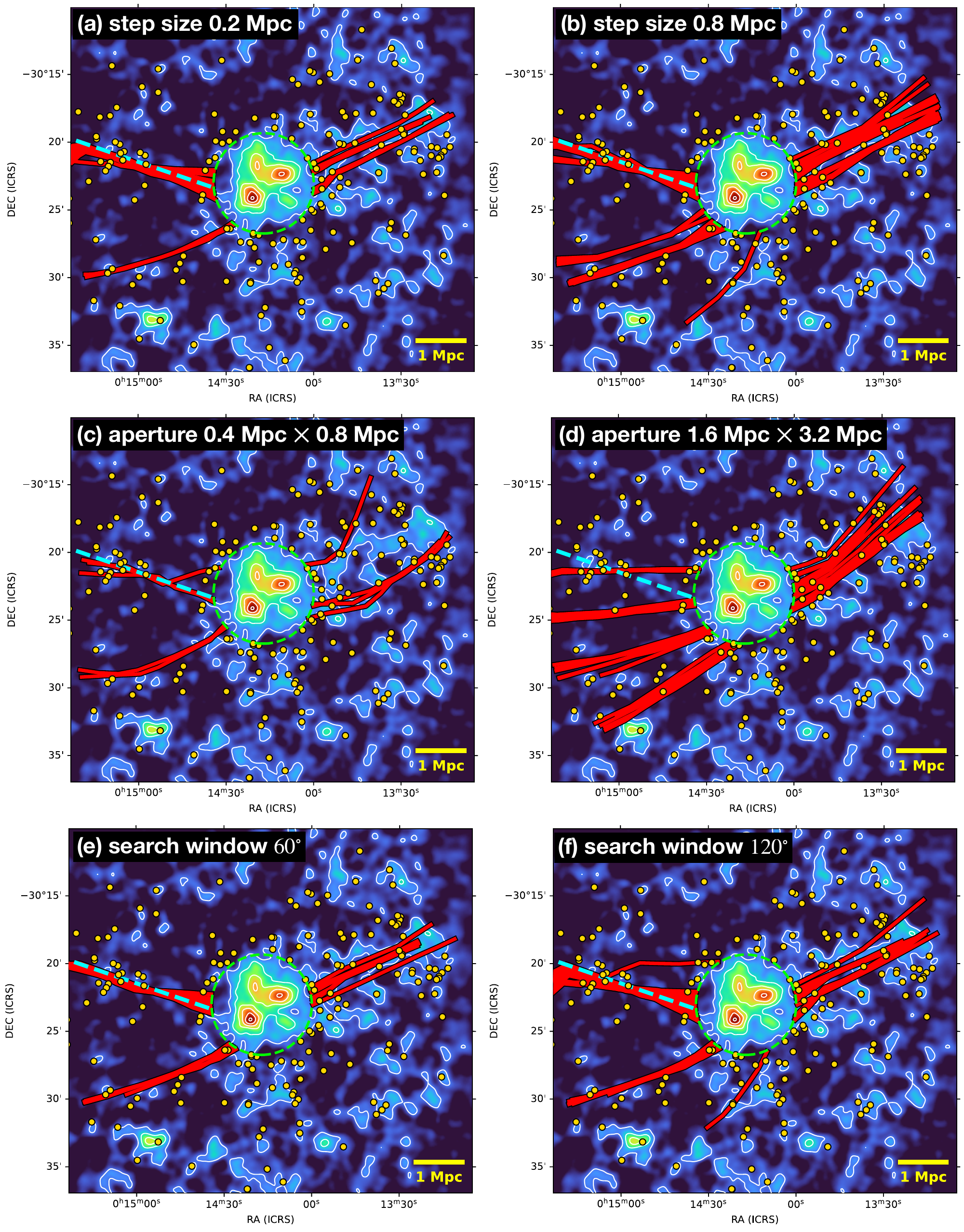}
\caption{Robustness test of the stepwise tracing method. The fiducial configuration in Figure~\ref{fig:2d_filament} uses an aperture with a width of $0.8 ~ {\rm Mpc}$ and a length of $1.6 ~{\rm Mpc}$, a step size of $0.4~ \rm Mpc$, and search window of $90^\circ$. Panels (a)–(f) show results when changing one setting at a time: varying the step size in (a) and (b), the aperture width and length in (c) and (d), and the search window in (e) and (f). In all panels, the yellow dots, the lime circle, the color map, and the contours are the same as in Figure~\ref{fig:2d_filament}(b). The dominant stream directions remain qualitatively consistent in (a)-(f), while (d) becomes less local and is therefore not adopted.}
\label{fig:step_by_step_robustness}
\end{figure*}

We perform a robustness test of the stepwise tracing method against the choice of aperture width, length, and step size. Our fiducial configuration uses an aperture with a width of $0.8 ~ {\rm Mpc}$ and a length of $1.6 ~{\rm Mpc}$, a step size of $0.4~\mathrm{Mpc}$, and a search window of $90^\circ$ (see \textsection\ref{method:2d_filament}). We repeat the reconstruction by changing the step size to $0.2~\mathrm{Mpc}$ and $0.8~\mathrm{Mpc}$ (Figure~\ref{fig:step_by_step_robustness}(a) and (b)), the aperture width and length to $0.4~\mathrm{Mpc}\times0.8~\mathrm{Mpc}$ and $1.6~\mathrm{Mpc}\times3.2~\mathrm{Mpc}$ (Figure~\ref{fig:step_by_step_robustness}(c) and (d)), and the search window of $60^\circ$ and $120^\circ$ (Figure~\ref{fig:step_by_step_robustness}(e) and (f)). Across these tests, the traced paths associated with the NW, E, and SE filament candidates
remain qualitatively similar to the fiducial result (Figure~\ref{fig:2d_filament}). Differences mainly appear in the local curvature and extent of the traced paths, as expected from the change in the sampling scale. For the largest aperture (Figure~\ref{fig:step_by_step_robustness}(d)), the tracing becomes less local, and the resulting paths are not directly comparable to the fiducial case.

\bibliography{main} 
\bibliographystyle{aasjournalv7}

\end{document}